\PassOptionsToPackage{dvipsnames}{xcolor}	
\documentclass[12pt,a4]{article}
\usepackage[T1]{fontenc}
\usepackage{lmodern}
\usepackage[utf8]{inputenc}
\usepackage{setspace}
\usepackage{etoolbox}
\makeatletter
\patchcmd{\appendix}{\@Alph}{\@Roman}{}{}
\makeatother

\usepackage{amsmath}
\usepackage{amssymb}
\usepackage{amsthm}
\usepackage{mathtools}
\usepackage{mathrsfs}
\usepackage{breqn}

\usepackage[margin=1in]{geometry}
\usepackage{enumerate}
\usepackage{enumitem}
\setlist[enumerate,1]{label=(\arabic*)}
\setlist[itemize,1]{label=--}    
\usepackage[dvipsnames]{xcolor}
\usepackage{hyperref}
\usepackage{float}
\usepackage{indentfirst}

\usepackage{sectsty}
\sectionfont{\centering}
\subsectionfont{\centering}

\usepackage{tikz}
\usetikzlibrary{decorations.pathreplacing}
\usepackage{mathtools}
\usetikzlibrary{positioning}
\usepackage{graphicx}

\newcommand{\bb}{\mathbb}

\newcommand{\U}{\bigcup}
\newcommand{\A}{\bigcap}
\newcommand{\und}{\underline}

\newcommand{\mcal}{\mathcal}

\newcommand{\supp}{\text{supp}}
\newcommand{\ext}{\text{ext}}

\renewcommand{\epsilon}{\varepsilon}

\newcommand{\blue}[1]{\color{blue}#1\color{black}}

\newtheorem{theorem}{Theorem}
\newtheorem{lemma}{Lemma}
\newtheorem{proposition}{Proposition}

\newtheorem*{statement*}{Result}

\theoremstyle{definition}
\newtheorem{definition}{Definition}

\DeclareMathOperator*{\argmax}{\arg\!\max}
\DeclareMathOperator*{\argmin}{\arg\!\min}
\DeclareTextFontCommand{\emph}{\slshape}

\usepackage{natbib}
\nocite{*}

\title{(Robust) Information Acquisition Design}

\author{Eric Gao\thanks{Department of Economics, Massachusetts Institute of Technology.  ericgao@mit.edu.} 
\space and Daniel Luo\thanks{Department of Economics, Massachusetts Institute of Technology. daniel57@mit.edu.} 
\footnote{Previous versions of this paper were circulated as ``(Prior-Free/Robust) Predictions for Persuasion'' and ``(Robust/Ex-Ante) Design of Persuasion Games.'' We are grateful to Itai Ashlagi, Ian Ball, Lucas Barros, B. Douglas Bernheim, Gonzalo Cisternas, 
Laura Doval, Drew Fudenberg, Kate Huang, Andrew Komo, Anna Merotto, Paul Milgrom, Ellen Muir, Axel Niemeyer, Alessandro Pavan, Parag Pathak, Harry Pei, Ryo Shirakawa, Eric Tang, Vitalii Tubdenov, Udayan Vaidya, Akhil Vohra, and especially Drew Fudenberg, Stephen Morris and Alex Wolitzky for helpful suggestions and feedback concerning this project. Refine.ink was used to check this manuscript for consistency and clarity. We are also thankful for helpful comments from participants at CalTech, MIT, Stanford, the 2023 MEA meetings, the 2023 Carroll Round, NASMES 2024, Stonybrook 2024, the Northwestern 2024 Summer School, the 2025 Asian Summer School in Economic Theory, and the 2025 ESWC Meetings. This paper was originally written when Gao and Luo were undergraduates at Stanford and Northwestern University respectively. Gao is grateful to the Stanford Department of Economics for financial support. Luo is grateful to the NSF Graduate Research Fellowship for financial support. All errors are, of course, our own.}}

\date{\today}

\begin{document}

\maketitle

\begin{abstract}
We study the design of information acquisition games---environments where a designer contracts their action on Sender's choice of experiment and the realized signals about some state. We develop a revelation-like principle for this setting and characterize the incentive properties of implementable allocations. We next turn to robust allocations---those that can be implemented regardless of the prior---and show that robustness is equivalent to experiment-neutrality of the mechanism. We conclude by considering two applications. First, for general good allocation problems, we show all efficient allocations are robust, even when agent preferences feature state-dependent outside options and allocation externalities. Second, we apply our model to school choice and uncover a novel informational justification for deferred acceptance when school preferences depend on students' unknown ability. 
\end{abstract}



\newpage
\onehalfspacing

\section{Introduction}
Decisionmakers often ask agents to acquire information before making a decision. For example, a school district may ask students to take exams before running a match; the FDA may ask a company to run a clinical trial before deciding whether or not to certify a new pharmaceutical; judges may ask for a period of discovery before hearing a trial. 
A critical feature of these interactions is that (1) the designer often commits to how they will respond to evidence before it is delivered (i.e. the school district writes the matching algorithm before students take their exams), and (2) transfers are prohibited. Despite the salience of these features in models of information acquisition and transmission, little is known about the economic content of these contracts.

In this paper, we build a general Sender-Designer model where Designer commits to an experiment-contingent action before Sender chooses a Blackwell experiment about a common, payoff-relevant state in order to analyze the above scenarios. Our first result, Theorem \ref{t: implementability}, characterizes the set of allocation rules which are implementable at a fixed prior, and thus speaks to the limits of economic contracting in this environment. In particular, we show it is without loss of generality to focus on truthful mechanisms (those that induce fully revealing experiments). Among all truthful mechanisms, we then show, via a Mirrleesian “shoot-the-sender” principle, that Sender’s binding deviation from truthful revelation must be the fully uninformative experiment. This allows us to verify implementability of an allocation rule for a fixed prior by checking a single inequality, which gives a simple test for whether an allocation rule is implementable in our model.

Having characterized implementable allocations, we next ask which allocations can be implemented \emph{robustly}, i.e. independently of the realized prior. Our exercise is motivated by \cite{bergemann2005robust}'s criticism of Bayesian mechanism design a la \cite{wilson1987game}, albeit in a very different context. Concretely, we imagine settings---particularly school choice and federal pharmaceutical regulation---where Designer must choose a uniform mechanism before interacting with many different agents, about each of whom Designer holds a different prior. In these settings, a useful property of the efficient allocation would then be that such an allocation can be implemented, regardless of the specific prior (i.e. which students apply to schools in a particular year). Our framework allows us to answer this question, and demonstrate its usefulness in applications.  

We give several properties of robust mechanisms. First, we prove a \emph{grim trigger robustness principle}, Theorem \ref{t: grim trigger robustness principle}: 
Robust mechanisms are exactly those which are \emph{experiment neutral}---they lead to the same action whenever Designer has the same posterior belief, regardless of which priors or experiments generated that posterior. 
Theorem \ref{t: grim trigger robustness principle} allows us to rewrite the problem of verifying robustness as a simple convex program, where whether an allocation rule is robustly implementable depends only on a finite set of linear inequalities (Proposition \ref{p: indifference beliefs}). Moreover, these binding beliefs depend only on Sender payoffs, and can be chosen independently of the allocation rule. This dramatically simplifies the problem of finding robust mechanisms by identifying, before writing down the allocation rule, a finite subset of incentive constraints which could bind. 

We next identify ordinal preference uncertainty as the core force which prevents certain allocation rules from being robust. First, we establish a tight anything-goes rule (Proposition \ref{p: homogeneous least favorites}): All allocation rules can be robustly implemented if and only if Sender has a state-independent least favorite action (i.e. an unequivocal way to ``shoot-the-sender''). We then specialize to the binary action case with single-crossing payoffs and propose a notion of ordinal monotonicity which completely characterizes robust allocation rules (Proposition \ref{p: monotone implementability}). These rules are exactly those which assign outcome $a$ with weakly higher probability at every state where Sender prefers $a$ than at any state where Sender prefers $b$. This mirrors monotonicity results in classical mechanism design, where cardinal monotonicity (i.e. the probability of being allocated the good is increasing in agents' types) is necessary.  

Finally, we study two applications. 
First, we interpret Designer as a school system designing a matching algorithm for students and schools, where school preferences vary with student ability.
The school system must announce a matching algorithm (Designer's contract) before students take exams, even though the specifics of the algorithm will affect student incentives to take exams (Sender experiments). Moreover, schools may interpret scores differently if students come from different backgrounds (prior-dependence), affecting the realized matching of students to schools (the allocation rule). Matches are robust in our setting exactly when students of the same ability are assigned to the same school, regardless of prior beliefs the school system may have about students. This desideratum---a version of (ex-ante) equal treatment of (ex-post) equal types (EAETEPE)---is inspired by recent policy interest in making admissions more ``meritocratic,'' i.e. a function only of realized student ability.\footnote{For example, the Supreme Court recently overturned affirmative action, at least partially based on the logic that students of equal ability were placed into different schools solely due to their race, where race may have influenced schools' prior beliefs about students' types (see \cite{SFFA_Harvard2023}). Similarly, the Boston school system was recently sued for switching three of its ``Exam Schools'' \emph{away} from deferred acceptance, with the plaintiffs alleging that this choice led to sorting orthogonal to realized ability (see \cite{boston_parent_cert}). Our results provide a theoretical channel through which movement away from deferred acceptance can violate an equal treatment of equals condition.}
Thus, a school system wishing to ensure their algorithm is prior-free not only needs to balance ex-post stability with interim incentives to report preferences over schools, but also student ex-ante incentives to take informative exams. Thus, when all three desiderata are considered, a modified version of \emph{student-proposing deferred acceptance} is the only algorithm which is robust and ex-post stable (interim strategy-proofness then comes for free). Our results suggest a novel reason deferred acceptance algorithms are used in practice---they shape informational incentives to ensure prior prejudices do not affect final matches. 

Second, we consider a government allocation problem where ex-ante uninformed agents (Senders) can commit to flexibly acquiring information\footnote{With many senders, we use \emph{type} instead of \emph{state} as each Sender designs information about some independent component of the information that only affects their (and Designer) payoffs.} after the government (Designer) chooses a mechanism. In this setting, we show the efficient allocation rule is always robustly implementable, contrary to standard negative results in the auctions literature when agent types are their private information (see \cite{JehielMoldovanu2001}). This result implies efficient implementation does not require Designer to restrict the prior beliefs Senders have about their type whenever Designer has commitment power in symmetric information environments, further shedding light on the driving force of negative results in the canonical mechanism design framework. 
Moreover, a particularly simple mechanism---a modified leave-one-out rule reminiscent of the Vickrey-Clarke-Groves mechanism (albeit without transfers)---robustly implements the efficient allocation rule. 

The remainder of this paper proceeds as follows. We next detail the related literature. Section 2 introduces the formal model. Section 3 analyzes the game for a fixed prior, and Section 4 characterizes robust mechanisms. Section 5 then considers the impact of ordinal preference uncertainty on implementability. Section 6 includes our applications. Section 7 concludes. Omitted proofs are collected in Appendix A, and Appendices B and C contain supplementary technical material. 

\subsection{Related Literature}
We draw from and contribute to three distinct strands of literature.

\vspace{-1em}
\paragraph{Contracting for Information Design.} 
Our paper relates to the literature on information design, initiated by \cite{KamenicaGentzkow11} and \cite{rayo2010optimal}, though we allow for Designer (Receiver) commitment. The closest paper to ours is thus  \cite{bergemann_gan_li}, who study a contracting environment where Designer commits to a decision rule before Sender chooses an experiment. Importantly, they fix the common prior and study robustness of Senders' set of potential experiments, in the spirit of \cite{Carroll2015}. In contrast, we (1) analyze the complete information game and establish the incentive constraints of implementation in that setting, and (2) fix the action set and vary the prior, focusing on prediction invariance instead of payoff guarantees. 
More generally, there is a recent literature in modifying persuasion to allow for settings where Designer commits to an allocation rule while Sender flexibly designs the common information environment. When experiments are costly, \cite{yoder2022designing} studies a binary-state model with transfers where Designer chooses an experiment and signal-contingent mechanism and characterizes optimal disclosure via a comparative statics result. \cite{yamashita2021bayesian} and \cite{terstiege2022competitive} study settings where Designer writes a contract in a transferable utility setting after an initial round of (public) Sender persuasion. When Designer can pay Sender, \cite{bergemann2002information} study how to implement efficient information acquisition. 
We contribute to this literature by studying a general model with Designer commitment before Sender information design, in the absence of transfers. We characterize the structure of (robust) allocation rules and the incentive constraints that underpin the limits of robustness.

\paragraph{Robust Mechanism Design.}
Our robustness exercise is motivated by \cite{bergemann2005robust} and \cite{bergemann2009robust}, who study mechanisms that implement the same allocation across type spaces, including agents’ beliefs and higher-order beliefs about payoff types; here we vary the common prior governing Sender’s information-acquisition problem.
Motivated also by the sensitivity of equilibrium predictions to the prior, \cite{andreyanov2021robust} study a general auction setting and characterize implementable outcomes that do not rely on any information about the prior; we see their motivation as complementary to ours for studying predictions robust to the prior. Further afield, \cite{oury2012continuous} study and characterize local robustness of mechanisms, while \cite{pei2024robust} study robust implementation when the designer must incentivize information acquisition about the state of the world. 
Using a different notion of robustness, but motivated by similar concerns about the fragility of predictions to unobservable belief-based approaches, \cite{dworczak_pavan_2022} study minimax-optimal information structures when Sender is worried Designer may have external information and seeks to guard their payoff against this information. 
Similarly, \cite{carroll2019robust} studies a minimax optimal design approach when there is uncertainty over which experiments are available to the agent. 
\cite{bergemann2013information} provides a detailed survey of the literature on mechanism design with endogenous information acquisition, both with and without robustness concerns. We contribute to this literature by introducing a new notion of robustness---to the \emph{prior} in games of information design---and characterizing its economic content in games of contracting for information acquisition. 

\paragraph{Auctions and Matching.} Our applications connect to two further literatures. In auction design, our efficient-implementation result stands in contrast to the impossibility theorems of \cite{JehielMoldovanu2001}, who show efficient design is generically impossible when agents hold private, interdependent valuations even in the prescence of transfers. The key distinction is that, in our setting, valuations are acquired rather than Sender's private information, and Designer commits before acquisition; we show this restores a possibility result, even without transfers. We thus relate to the literature on allocative externalities, studied by \cite{jehiel1996how}, \cite{dasvarma_2002_standard}, \cite{jehiel2006allocative}, and \cite{kuehn_2019_estimating}. In matching, we build on the stability and strategy-proofness foundations of \cite{gale_1962_college} and \cite{roth_1982_the}, \cite{roth_1989_twosided}. We also relate to work on stable matching under incomplete information, including \cite{liu_2014_stable}, \cite{bikhchandani_2017_stability}, and \cite{fernandez_2021_centralized}. Whereas that literature largely asks how to define stability when preferences are uncertain, we fix ex-post stability and ask which algorithms robustly implement it when school priorities depend on unobserved ability; our proof leverages the strategy-proofness-exposing description of deferred acceptance due to \cite{yannai}. The resulting essential uniqueness of student-proposing deferred acceptance—which excludes Designer-proposing and compensation-chain algorithms \cite{dworczak}---provides a novel informational justification for its use.

\section{Model}
There exists a \emph{Designer} who takes one of finitely many actions $a \in \mcal A$, and one\footnote{We emphasize here that our results extend to the case where there are arbitrarily many Senders, each of whom observe part of the state, drawn independently (as is the case in our applications). However, for notational and expositional simplicity, we present our characterization for a single Sender in the main body, and leave the extension to many senders in \blue{\ref{Appendix B}}.}  \emph{Sender} who can acquire information about some state drawn from a finite set $\mcal T$ according to a common full-support prior distribution $\mu^0$.
Sender's Bernoulli payoffs are given by $u(a, t)$.

Strategies are as follows. Let $S$ be an arbitrary (but fixed throughout) Borel space. First, Designer commits to a (Borel measurable) mechanism $m: \Delta(S)^{\mcal T} \times \Delta(\mcal T) \to \Delta(A)$, which outputs a random action as a function of the experiment Sender chooses and the realized posterior. After observing $m$, Sender chooses a Blackwell experiment $\mathscr{E}: \mcal T \to \Delta(S)$. When there is no ambiguity (e.g whenever the prior is fixed), we will interchange between the experiment-based approach and the belief-based approach\footnote{See \cite{KamenicaGentzkow11} and \cite{bergemann2016bayes}. We will refer to this equivalence whenever there is a fixed prior belief, and otherwise rely on the experiment-based formulation.}, in which case Blackwell experiments are Bayes-plausible distributions of posterior beliefs $\tau \in \Delta(\Delta(\mcal T))$ such that $\bb{E}_\tau[\mu] = \mu_0$, while mechanisms are functions $m: \Delta_{\mu_0}(\Delta(\mcal T)) \times \Delta(\mcal T) \to \Delta(A)$, where $\Delta_{\mu_0}(\Delta(\mcal T))$ is the set of Bayes-plausible distributions at prior $\mu_0$. 

Given a mechanism $m$ and experiment $\tau$, expected payoffs for the Sender can be represented as 
\[ \mcal U(\tau, m) = \int_{\Delta(\mcal T)} \int_{\mcal T} \int_{\mcal A} u(a, t) dm(\tau, \mu) d\mu d\tau.  \]
If, given some mechanism $m$ and prior $\mu^0$, Sender's strategy is a best response to the mechanism, we say that $m$ \emph{implements} the strategy $\tau$. That is, $m$ \emph{implements} $\tau$ if 
\[ \tau \in \argmax\left\{\mcal U(\tau', m) : \int \mu d\tau' = \mu^0\right\}. \]
Let $\Sigma(m, \mu^0) \subset \Delta(\Delta(\mcal T))$ be the set of all strategies implemented by $m$ at prior $\mu^0$. Of course, there is also a corresponding set of experiments $\Sigma^{\mathscr{E}}(m, \mu^0) \subset \Delta(S)^{\mcal T}$ that exists. 

Given any mechanism $m$ and strategy $\tau \in \Sigma(m, \mu^0)$, the induced \emph{allocation rule} is the function $p_{m(\tau, \cdot), \tau}(t) = \int m_\tau(\mu) d\tau(\mu | t)$, tracking the distribution of actions when a specific state $t$ is realized. When $p(t)$ is the degenerate lottery over one element, we will abuse notation and set $p(t) = a$ to refer to type $t$ being assigned a degenerate lottery at action $a$.
We interpret the allocation rule as the ex-post data Designer has access to, noting it governs Sender's state-by-state realized payoffs. 
When $p_{m(\tau, \cdot), \tau}$ is induced by $m(\tau)$, we can represent ex-ante payoffs as 
\[ \int u(p(t), t) d\mu^0 = \mcal U(\tau, m) = \int u(p_{m(\tau, \cdot), \tau)}, t) d\mu^0(t). \]
Here we abuse notation and set $u(p(t), t) = \bb{E}_{m, \tau}[u(a, t) | t]$. 
\begin{definition}
    \label{d: implementation}
    A mechanism $m$ implements allocation rule $p$ at prior $\mu^0$ if there exists strategy $\tau \in \Sigma(m, \mu^0)$ such that $p = p_{m(\tau, \cdot), \tau)}$ at $(m, \tau)$. Equivalently, Figure \ref{arrowsgobrr} commutes.
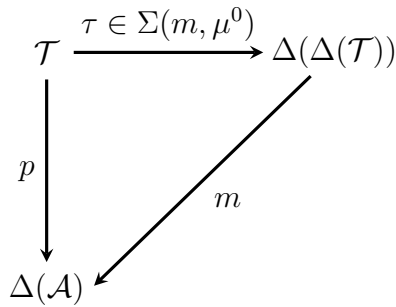
\begin{figure}[h]
	\centering
	\begin{tikzpicture}[scale=1]
		\node[] at (0,0) {$\mathcal{T}$};
		\node[] at (4.25,0) {$\Delta(\Delta(\mathcal{T}))$};
		\node[] at (0,-3.5) {$\Delta(\mcal A)$};
		
		\draw[black, very thick,-stealth](0.4,0)-- node[above] {$\tau \in \Sigma(m, \mu^0)$} ++(2.8,0);
		\draw[black, very thick,-stealth](0,-0.4)-- node[left] {$p$} ++(0,-2.7);
		\draw[black, very thick,-stealth](3.9,-0.35)-- node[below right] {$m$} ++(-3.2,-3.1);
	\end{tikzpicture}
    \caption{Diagrammatic Representation of Implementability}
    \label{arrowsgobrr}
\end{figure}

If there exists $m$ that implements $p$ at $\mu^0$, then $p$ is \emph{implementable} at $\mu^0$. 
\end{definition}


\section{Implementability}
We first characterize the set of implementable allocations given a prior $\mu^0$. To do this, we will require the following definitions. Throughout this section, fix a full-support prior $\mu^0$. 

\begin{definition}
    \label{d: direct}
    A mechanism $m$ is direct if there exists $\tau \in \Sigma(m, \mu^0)$ such that $\tau$ is \emph{fully revealing}: it supports only degenerate beliefs. 
\end{definition}

\begin{definition}
      For any belief $\mu$, the \emph{grim-trigger correspondence} $r(\mu): \Delta(\mcal T) \rightrightarrows \Delta(\mcal A)$ is defined by
    \[ r(\mu) = \argmin_{a \in \mcal A} \int u(a, t) d\mu. \] 
\end{definition}

When there is no ambiguity, we will abuse notation and refer to $r(\mu)$ both as a correspondence and a selection from that correspondence, for example $u(r(\mu), t)$ denotes Sender's payoff from the grim-trigger belief at $\mu$, given type $t$.

\begin{theorem}[Grim-Trigger Revelation Principle]
    \label{t: implementability}
    The following are equivalent. 
    \begin{enumerate}
        \item There exists a mechanism $m$ which implements $p$ at prior $\mu^0$. 
        \item There exists a direct mechanism $m$ which implements $p$ at prior $\mu^0$. 
        \item The following inequality holds.
        \[ \int u(p(t), t) d\mu^0 \geq \int u(r(\mu^0), t) d\mu^0\]
    \end{enumerate}
\end{theorem}

Theorem \ref{t: implementability} has two parts. The first part is analogous to the standard revelation principle: there are \emph{direct} mechanisms that implement any implementable allocation. Our direct mechanisms are distinct from the standard Myersonian notion: We say that a mechanism is direct (i.e. ``truthful'') if it is fully revealing. That is, the designer has no ambiguity about Sender's type after seeing the realization of the sender's signal. This differs from the standard setting where types are agents' private information and mechanisms are direct if agents truthfully report, since we work in a symmetric information environment. 

The second part is the equivalence between implementation and an incentive compatibility constraint. Since Sender has no private information, they do not collect any information rents. Thus, one potential intuition is that the mechanism design problem is “free:” any allocation can be implemented without cost. However, despite not collecting information rents, Sender still collects \emph{incentive} rents since Sender ultimately has the power to choose the information structure. In particular, Theorem \ref{t: implementability} shows that Designer must still respect \emph{average} incentive compatibility, where types are averages relative to their weight by the prior. This is because, if the allocation is on average bad for types, than Sender can always choose the no-information experiment, and collect the exogenous outside option given by $\int u(r(\mu^0), t) d\mu^0$. 

The intuition behind the proof of Theorem \ref{t: implementability} is as follows. First, it is clear that (2) implies (1). To show that (1) implies (2), let $m$ implement $p$ via $\tau$. Consider the following ``shoot-the-sender'' mechanism: give Sender $p(t)$ (type-by-type) if Sender's chosen experiment is fully revealing\footnote{Indeed, conditioning on the experiment is crucial for the first equivalence to hold. In \blue{\ref{Appendix C}}, we give an example where belief-based mechanisms cannot trace out the set of all implementable allocations.}. Otherwise, at every realized belief $\mu$ under $\tau$, give Sender a selection from the grim-trigger correspondence, $r(\mu)$. Since  $u(p(t), t)$ is incentive compatible under the original mechanism, and Sender cannot do worse than get $r(\mu)$ at any belief, this new ``shoot-the-sender''' mechanism will inherit incentive compatibility. The details are spelled out in \blue{\ref{proof of t: implementability}} 

That (2) implies (3) follows from the fact that if $m$ implements $p$, then in particular it must ward off the fully uninformative experiment, which gives Sender a payoff of at least $\int u(r(\mu^0), t) d\mu^0$. That (3) implies (2) follows by applying Blackwell's theorem to the allocation: at any nondegenerate experiment, Sender is receiving $r(\mu)$, and so they do better by providing Blackwell less information. Since no-information is Blackwell minimal, it must be the most profitable deviation. Thus, warding off no information implies that all other deviations are unprofitable as well. As before, the details are spelled out in \blue{\ref{proof of t: implementability}} 

\section{Robustness}
\subsection{Robust Implementation}
Theorem \ref{t: implementability} characterizes the incentive costs of motivating truthful revelation when the prior is known to the principal and agent. In this section, we ask a related question: what is the incentive cost of motivating truthful revelation \emph{regardless} of the prior? 

Our extension of Theorem \ref{t: implementability} is motivated by a long-run designer's objective when setting up a long-term mechanism without knowing the prior, which is then used in various short-term instances with varying, known priors\footnote{We see this exercise as being similar in spirit to that of \cite{bergemann2005robust}, but for our contracting environment.}.
Such is the case with our school-choice application: the designer must design a \emph{single} mechanism that is effective for \emph{all students}, regardless of the prior belief that schools have about individual students. 
To always implement a desirable allocation in settings like this requires a \emph{robust} mechanism: one for which implementation is prior-free. This leads us to the following definition.

\begin{definition}
    \label{d: robust definition}
    A mechanism $m$ is \emph{robust} if there exists $p$ such that $m$ implements $p$ for all full-support priors $\mu^0$. An allocation rule $p$ is \emph{robustly implementable} (or just \emph{robust}) if there exists a robust mechanism $m$ implementing $p$ for all full-support priors $\mu^0$. 
\end{definition}

Definition \ref{d: robust definition} thus captures this section's motivating question: an allocation rule $p$ is robustly implementable if and only if its implementability does not depend on the prior. 

\subsection{The Grim-Trigger Robustness Principle}
Theorem \ref{t: implementability} gives a complete, prior-by-prior characterization of implementation, but notably condition (3) requires prior-dependence and hence it is unable to speak to robust allocation rules. In general, characterizing robustness may be more difficult since it requires the induced allocation rule $p(t)$ remain invariant to both changes in the prior likelihood of each state and the induced second-order change in the set of Sender’s feasible posterior distributions through Bayes plausibility. For general mechanisms, these two forces can interact in potentially complex
ways. We simplify the problem by identifying a smaller set of mechanisms we can restrict to
without loss of generality; these lead to a simple, complete characterization of robustness.

\begin{definition}
    A mechanism $m$ is \emph{belief-based} if, for any tuples $(\tau, \mu)$ and $(\tau', \mu)$, $m(\tau, \mu) = m(\tau', \mu)$. $m$ is a \emph{grim-trigger} mechanism if, whenever $\mu$ is nondegenerate, $m(\tau, \mu) \in r(\mu)$. 
\end{definition}

Belief-based mechanisms require Designer's mechanism to be independent of the choice of experiment.\footnote{Interestingly, a revelation principle need not hold among all belief-based mechanisms, as the implemented allocation rule may randomize in complex ways. We discuss in detail the richness in indirect belief-based implementation in \blue{\ref{Appendix C}}. Note our result in this setting contrasts with Theorem 3 in \cite{yoder2022designing}.} Whenever $m$ is belief-based, we will drop the dependence on $\tau$ and only write $m(\mu)$. Because belief-based mechanisms are independent of the experiment, they are also implicitly independent of the prior belief.
As a result, belief-based mechanisms obviate the complicated interaction between the prior and the feasible set of experiments and thus are a natural candidate for robust mechanisms. 

Our main result states that (1) an allocation is robust if and only if it has a direct, belief-based grim trigger implementation, and (2) whether or not an allocation rule can be implemented by a robust mechanism depends solely on how Sender's expected utility at nondegenerate beliefs compares against the grim-trigger payoff at those beliefs.  

\begin{theorem}[Grim-Trigger Robustness Principle]
\label{t: grim trigger robustness principle}
The following are equivalent. 
\begin{enumerate}
    \item (Robustness). An allocation rule $p$ is robustly implementable. 
    \item (Experiment-Neutrality). $p$ is implementable by a direct, belief-based, grim-trigger mechanism at some full-support prior. 
    \item (Pointwise Incentive Compatibility). For all beliefs $\mu \in \Delta(\mcal T)$, 
    \[ \int u(p(t), t) d\mu \geq \int u(r(\mu), t) d\mu. \]
\end{enumerate}
\end{theorem}

Theorem \ref{t: grim trigger robustness principle} has three implications. 
First, whenever (and only when) an allocation $p$ can be implemented by a direct, belief-based, grim-trigger mechanism at \emph{some} prior, it is robust: it is implementable at all prior beliefs. That is, \emph{robustness} is equivalent to \emph{experiment-neutrality}: any prediction which does not condition on the prior can be implemented by a mechanism which ignores the choice of experiments (i.e. counterfactual inferences that may have been drawn from other signal realizations) and which only depends on the experiment through its realized posterior beliefs.
We emphasize here that the equivalence between (1) and (2) is not immediate: prior freeness a-priori does not require experiment neutrality, since it is in principle possible to condition on the entire distribution of posterior beliefs while still maintaining independence to the prior (for example, the ``shoot-the-sender'' mechanisms of Theorem \ref{t: implementability} are in principle prior-free). Theorem \ref{t: grim trigger robustness principle}, however, shows that so long as an allocation is robustly implementable, one can \emph{find} a belief-based mechanism which implements it. Indeed finding such a mechanism is subtle, and the proof of the equivalence between (1) and (2) is not constructive; we remark on this below. 

Second, robust implementation of $p$ boils down to the (simpler) problem of whether we can find some prior $\mu^0$ at which $p$ can be implemented by a direct, belief-based mechanism, an easier condition to verify than the universal quantifier in the definition of robustness. We remark the change in the order of quantifiers is not obvious; a natural guess for (2) would be that $p$ is implementable by a direct, belief-based, grim-trigger mechanism at \emph{every} prior. The gap, then, comes in showing that experiment-neutrality implies prior-freeness: and so the existence of such a mechanism for some prior translates to the existence of a robust mechanism. 

Third, the third part of the equivalence is best contrasted with the third part of the equivalence of Theorem \ref{t: implementability}. In particular, instead of requiring implementability at the prior belief, robust implementation requires implementability at \emph{every} belief. That this is necessary is clear; the potentially surprising part is that this condition is also sufficient: implementation at each belief is equivalent to being able to find a prior-independent mechanism implementing this allocation. The cost of robustness is thus straightforward: it is exactly the cost of needing to implement the allocation, even as beliefs vary arbitrarily. 

The intuition behind the argument is as follows. First, that (1) and (3) are equivalent is straightforward, and follows by strengthening the prior-by-prior characterization in Theorem \ref{t: implementability}. That (2) and (3) are equivalent is somewhat more subtle. The key step behind the equivalence is recognizing that belief-based mechanisms differ from the ``shoot-the-sender'' mechanisms of Theorem \ref{t: implementability} in that they cannot assign the grim-trigger outcome $r(\cdot)$ at degenerate beliefs which are part of a partially pooling experiment $\tau$, and instead must assign $p(t)$. Thus, deviations where Sender induces a profitable belief $\mu$ violating (3) with some probability while revealing the state with complementary probability become particularly profitable: they are guaranteed $p(t)$ when revealing the state, but can obtain 
\[ \int u(r(\mu), t) d\mu \geq \int u(p(t), t) d\mu\]
at any belief $\mu$ where the allocation $p(t)$ is not profitable. To ward off against deviations of this form then, pointwise incentive constraints must hold exactly for every belief $\mu$. 
The details for both equivalences can be found in \blue{\ref{proof of t: grim trigger robustness principle}}

\subsection{Belief-Based Incentive Constraints}
The third equivalence in Theorem \ref{t: grim trigger robustness principle} identifies a set of belief-by-belief incentive constraints which pin down whether an allocation rule is robustly implementable. This significantly reduces the dimensionality of the problem, where one originally needed to show that fully revealing experiments were more profitable for Sender than any possible {distribution} of beliefs. 
However, the current program is still complicated in two ways. First, there remain a continuum of incentive constraints that must be verified to determine if an allocation rule can be robustly implemented. Second, the specific beliefs at which incentive constraints may bind can vary with the choice of allocation rule, making it difficult to characterize the {entire} set of robustly implementable allocation rules.  

In this section, we exploit the {convex} structure of incentive constraints in Theorem \ref{t: grim trigger robustness principle} to rectify both of these issues by identifying a {finite} set of beliefs which pins down robustness, and show that these beliefs can be chosen independent of the allocation rule whose robust implementability we wish to verify. Towards this end, we employ the following definition. 

\begin{definition}
\label{d: indifference beliefs}
 Let \emph{indifference beliefs} $\mcal D = \left\{ \mu : |r(\mu)| \geq 2 \right\}$ be the set of beliefs where the grim-trigger correspondence is multiple-valued. 
\end{definition}

Importantly, the set of indifference beliefs can be determined independently of the mechanism or allocation rule. Moreover, it has a convex structure: The sure-thing principle implies that if $a \in r(\mu)$ and $a \in r(\mu')$  then $a \in r(\beta \mu + (1 - \beta)\mu')$ for any $\beta \in [0, 1]$. Thus, the sets $K(a) = \{\mu : a \in r(\mu)\}$ are convex and $\mcal D$ is (generically, in the absence of ties) exactly the set of all boundary points of these sets which are interior to $\Delta(\mcal T)$. 

			
			
			
			

It turns out that extremal indifference beliefs are exactly those which pin down the incentive constraints for robust implementability. Formally, 

\begin{proposition}[Finite IC]
\label{p: indifference beliefs}
Define $\mcal E = \U_{a \in \mcal A} \text{ext}(K(a)) \subset \ext(\Delta(\mcal T)) \cup \mcal D$ to be a finite set of degenerate or indifference beliefs. 
The following values are equivalent:  
\begin{align*}
\min_{\mu \in \Delta(\mcal T)} \left\{ \int u(p(t), t) - u(r(\mu), t) d\mu \right\} = \min_{\mu \in \mcal E} \left\{ \int u(p(t), t) - u(r(\mu), t) d\mu \right\}
\end{align*} 
Thus, $p$ is robust if and only if $p$ satisfies pointwise incentive constraints on $\mcal E$. 
\end{proposition}

The basic idea behind the proof is to apply Bauer's theorem of the maximum on the sets $K(a)$ and then show $K(a)$ is a finite polytope, which establishes that $\mcal E$ is finite. To moreover show that $\mcal E$ contains only indifference and degenerate beliefs follows from the fact that these are exactly the points on the boundaries of the $K(a)$, though the formal proof requires some topological quibbling. The details are ironed out in \blue{\ref{proof of p: indifference beliefs}.}

\section{(Robust) Ordinal Preference Uncertainty}

\subsection{Homogeneous Least Favorites}

What are the limits of robust contracting? Consider the case where Sender has state-independent preferences, so that $u(a, t) = u(a)$. Then there exists some $a_{LF}$ such that $u(a) \geq u(a_{LF})$ for all $a$, and hence the grim-trigger correspondence has a constant (worst) selection. When this is the case, pointwise incentive constraints are trivially satisfied for every allocation rule and all direct, belief-based, grim-trigger mechanisms are robust (and hence every allocation rule can be robustly implementable). This then implies that, for example, every outcome is robust in the preference specification of \cite{KamenicaGentzkow11}'s judge-prosecutor motivating example: Any outcome can be rationalized if the judge has commitment without needing to take a stance on the prior. 

The above discussion suggests preference heterogeneity among Sender' least favorite actions at different states is one key force that may require one to take a stance on the prior players in the game have. We show this connection is tight: Every allocation rule is robust if and only if Sender has the same least favorite action at each state. Formally, 

\begin{definition}
\label{d: homogenous least favorites}
Sender preferences satisfy \emph{homogeneous least favorites} if $r(\mu)$ has a constant selection in $\mu$.  
\end{definition}

\begin{proposition}[Homogeneous Least Favorites]
\label{p: homogeneous least favorites}
Every allocation rule $p$ is robust if and only if Sender preferences satisfy homogeneous least favorites. 

\end{proposition}

The converse is immediate. To show the forward direction, we construct a non-constant allocation rule which gives Receiver their least favorite at every type, and show that pooling onto the uninformative experiment must be more profitable since it can give at most one action, which is (strictly) better than receiving the type-dependent least favorite action. 

Our results are consistent with other literature on robust contracting in persuasion settings, namely \cite{bergemann_gan_li}. There, if Sender's preferences are state-independent (a stronger notion than homogeneous least favorites), Designer can also implement any allocation rule. Crucially, though, their results fix a prior, vary ambiguity over the space of experiments, and rely on Sender's choice of tiebreaking rule. 
The contrapositive of our result is particularly useful, as there are many settings (such as our applications) where the homogeneous least favorites condition is likely to be violated. In school-choice, students' preferences can vary significantly with their ability, especially among specialist schools (i.e. musical conservatories versus elite academic schools). In auctions, allocation externalities can cause variation in whether some sender prefers a good to be thrown away versus given to a different sender. We characterize robust allocation rules in these settings in the applications. 



\subsection{Binary Actions and Ordinal Monotonicity}
Throughout this section, suppose there are only two actions $\{a_1, a_2\}$. It must be that any indifference belief $\mu \in \mcal E$ exactly makes Sender indifferent between $a_1$ and $a_2$. This then implies that pointwise incentive constraints at some $\mu \in \mcal E$ take the form 
\[ \int (u(a_1, t) - u(a_2, t)) \bb{P}(p(t) = a_1) d\mu(t) \geq 0.\]
This simple structure of pointwise incentive compatibility constraints will allow us to say much more about robust implementation. We will need the following definition. 

\begin{definition}
   $p: \mcal T \to \Delta(\mcal A)$ is \emph{ordinally monotone} if 
   \[  \inf_{t: u(a_1, t) \geq u(a_2, t)}\bb{P}(p(t) = a_1) \geq \sup_{t: u(a_1, t) < u(a_2, t)}\bb{P}(p(t) = a_1). \]
\end{definition}

An allocation rule is ordinally monotone if and only if it assigns ``higher'' states the ``good'' outcome with higher probability. When $a_1$ is interpreted as an agent being allocated a good and $\mcal T$ is their valuation, this definition exactly coincides with standard notions of monotonicity in classical mechanism design. However, our setting differs in two important ways. First, Sender valuations are not perfectly observed, and instead Sender must experiment to reveal (and form inferences) about the state. Second, Designer cannot use transfers. Despite this, we are able to obtain a (modified) version of the celebrated result that all (and only) monotone allocation rules are implementable.\footnote{See \cite{sinander} for a detailed reference and discussion in the classical mechanism design setting.} The proof is in \blue{\ref{proof of p: monotone implementability}}

\begin{proposition}
\label{p: monotone implementability}
$p$ is robustly implementable if and only if it is ordinally monotone. 
\end{proposition}

Proposition \ref{p: monotone implementability} states an allocation rule is robust if and only if $a_1$ is allocated with higher probability at all states whom prefer $a_1$ to $a_2$ then at any state where $a_2$ is more preferred. 
The intuition is that, because Designer can always discipline revelation among states with the same ordinal preferences by threatening to give them the other outcome, (robust) implementation is constrained exactly (and only) by how the allocation rule distinguishes between states where preferences flip. 

In \blue{\ref{Appendix B.I}}, we generalize to the case where there are many players, and show that so long as interim payoffs for each Sender behave ``as-if'' there are two goods,\footnote{This is satisfied in good allocation rule problems, for example, where each Sender only cares whether or not they were allocated the good.} an interim version of Proposition \ref{p: monotone implementability} holds. This extension allows us to obtain sharp results about robust implementation in good allocation rule settings with allocation externalities. We apply our formalism to study the case with allocation externalities in the applications. 

\section{Applications}
Throughout our applications, we consider the case in which there is more than one sender. In these cases, our notions of implementation and robustness hold in the interim (supposing other players choose fully revealing experiments). Robustness then requires that interim payoffs are weakly greater than the grim trigger payoff, regardless of the prior belief that is held. Because the extension is mostly technical, and develops no new ideas, we relegate it to \blue{\ref{Appendix B}.} \quad All relevant theorems, however, extend exactly as one expects. 

\subsection{Allocating a Good with Externalities}

We study the problem of good allocation when Designer must incentivize agents to acquire information about their valuations and transfers are not feasible as a way to screen. 
When all agents strictly prefer being allocated the good regardless of their type, Proposition \ref{p: homogeneous least favorites} implies every allocation rule is robust. However, this need not be true when Senders have type-dependent outside options\footnote{For example, if the experiment is informative about each sender's outside options.} or allocation externalities.
When this is the case, there is disagreement about each Sender's least favorite option, and thus robustness of allocation rules is not obvious (for example, a rule which gives the good to whichever Sender values it the least is not robust).

Despite this, we show in this section that the efficient allocation rule is robust in good allocation problems with very general preferences. We interpret this proposition as a possibility result that highlights the irrelevance of specifying a prior belief for the efficient design of information acquisition for general good allocation problems. This contrasts, for example, with results in the private-information setting (see \cite{bergemann2005robust}). 

Formally, suppose there are $\mcal I$ senders and the set of actions $\mcal A = \mcal I \cup \varnothing$ specifies which agent is allocated some good (with the option to throw away the good if Designer wants, though doing so gives the designer no value.). Payoffs are given by $u_i: \mcal A \times \mcal T_i \to \bb{R}$. 
We require only the following assumption on preferences, which implies that no player benefits from the good being allocated to someone else (that is, each agent would prefer the good to be thrown out rather than allocated to some competitor). This assumption is satisfied for any type-dependent outside options and all applications in the examples cited in the previous literature, and is reminiscent to the negative externalities assumption of \cite{jehiel2006allocative}. 

\begin{definition}
    Payoffs $u_i: \mcal A \times \mcal T_i \to \bb{R}$ satisfy \emph{weakly negative externalities} if $u_i(j, t_i) \leq 0$ for all $j \neq i$ and $t_i \in \mcal T_i$, while $u_i(i, t_i) \geq 0$ for all $i, t_i \in \mcal T_i$. 
\end{definition}
Define the efficient allocation to be 
\[ p^*(t) = \argmax_{a \in \mcal A} \left\{ \sum_{i \in \mcal I} u_i(a, t_i) \right\}. \]
where we take any selection if $p^*(t)$ is multiple-valued. Note that by scaling payoffs, we can vary $p^*(t)$ to be any supported Pareto-efficient allocation.
Finally, define the \emph{leave-one-out} mechanism to be the direct, belief-based mechanism which allocates 
\[ p_{-i}(t_{-i}) \in \argmax_{k \in \mcal A \setminus \{i\}} \left\{\sum_{j \in \mcal I \setminus \{i\}} u_j(k, t_j) \right\} \]
when only agent $i$ unilaterally deviates.\footnote{See \blue{\ref{Appendix B.I} } for a formal definition of direct, belief-based mechanisms for many senders.} We have the following result, whose proof can be found in \blue{\ref{proof of p: efficient allocation}}.

\begin{proposition}
\label{p: efficient allocation}
    Suppose payoffs satisfy weakly negative externalities. Then the efficient allocation rule $p^*(t)$ is robustly implemented by the leave-one-out mechanism. 
\end{proposition}

Proposition \ref{p: efficient allocation} implies not only that efficient allocation rules are robust, but also that they robustly implemented by a particularly tractable class of mechanisms---those that, instead of minimizing the payoffs of unilateral deviators (a particularly harsh punishment) simply ``rejects'' those deviators and commits to the efficient allocation rule among all other agents. 

The core intuition behind the result has two parts. First, efficient implementation requires maximizing aggregate surplus, and hence there are no conflicts of interest towards good allocation because Designer can commit to disposing of the good when necessary. Second, because all externalities are weakly negative, the leave-one-out mechanism guarantees a weakly lower payoff to each Sender than if they commit to the truthful mechanism, echoing the canonical intuition of Vickrey-Clarke-Groves types mechanisms in auction design. Consequently, the efficient allocation rule (with leave-one-out punishments) exactly ordinally aligns Sender incentives posterior-by-posterior, which is sufficient for robust implementation. 

\subsection{Stable Matching}
Consider the problem faced by some centralized school system, which must allocate thousands of students to heterogeneous schools with different characteristics.
Districts often must write their matching algorithm before students take exams or report preferences.\footnote{A plethora of centralized admissions systems set a matching algorithm beforehand and do not take into account the effects this has on students' incentives to acquire information about their ability; New York, Chicago, and Boston all implement centralized matching algorithms for high-schools; Taiwan, China, Korea, and others run similar systems for college, and the residency match for doctors in America famously employs a deferred acceptance algorithm.} If school priorities over students depend on students' unobserved abilities (which are statistically revealed through exams), the matching algorithm may affect student behavior leading up to exams,\footnote{For example, how long to study, which exams to take, how many classes to register for, etc.} and thus affect realized school priorities and the final match. 
Because school inferences about student ability may vary with their prior belief about the student before seeing the exam score (i.e. parental occupation, residential code, race, etc.), different matching algorithms may cause students of equal ability to be matched to different schools, based only on differences in the school system's prior belief and not their realized exam scores, violating \emph{ex-ante equal treatment of ex-post equals} (EAETEPE).

Violating such a desiderata may be politically infeasible or unpopular, and so a designer may wish to avoid matching algorithms which engender unequal treatment, especially in the wake of the decision in \emph{Students for Fair Admissions vs. Harvard}.\footnote{The Supreme Court explicitly writes ``the Court has permitted race-based college admissions only within the confines of narrow restrictions: such admissions programs must comply with strict scrutiny, may never use race as a stereotype or negative, and must—at some point—end'' \citep{SFFA_Harvard2023}. Similarly, in \emph{Bakke vs. Regents of the University of California}, the Court writes ``The guarantee of equal protection cannot mean one thing when applied to one individual and something else when applied to a person of another color'' \citep{regents1978}. We interpret this criteria as a form of EAETEPE: students of different races cannot be given different outcomes if they have the same ability.} For example, the Boston school system was sued when they moved away from student-proposing deferred acceptance for elite schools, with families alleging that doing so would violate equal treatment of students with equal ability, but different prior characteristics (see \cite{boston_parent_cert}). 

We use our framework to understand which matching algorithms satisfy anti-discrimination standards, i.e. satisfy an EAETEPE-like criterion. To prevent us from deviating too far from canonical models, we also impose traditional desiderata---particularly ex-post stability---on the algorithm.
We show a modified, ``test, then ask'' version of deferred acceptance, which collapses to standard deferred acceptance when ability is observed (or when school priorities are independent of student ability) is essentially the only algorithm which robustly implements a stable outcome. We interpret this as an additional \emph{equitable information} justification for student-proposing deferred acceptance in school choice.

Formally, fix a two-sided one-to-one matching market with $i \in \mcal I$ senders and $j \in \mcal J$ schools, with $|\mcal I|=|\mcal J|$. The set of actions $\mcal A$ are possible matches.\footnote{Formally a function $a: \mcal I \cup \mcal J \to \mcal I \cup \mcal J$ where $a(j) \in \mcal I, a(i) \in \mcal J$, and $a(i) = j \iff a(j) = i$ for all $i, j$.}
Students' preferences $u_i(j, t_i)$ depend on the school they are matched with and their idiosyncratic type, so $u_i(a, t_i) = u_i(a(i), t_i)$. Suppose student preferences are strict.
Schools have preferences $v_j(a, t) = v_j(a(j), t_{a(j)})$ which depend on the specific student they are matched with and the type\footnote{Each sender's type $t_i$ can be interpreted as their ``state'' in the corresponding interim version of the problem. We use type in both applications to clarify and draw connections to the standard mechanism design literature, and highlight that each type does not affect other senders' payoffs or informational environment.} of that student\footnote{School preferences do not play a role in the analysis and are only included in order to contrast our model with the standard matching model.}. 
Designer commits to a matching algorithm given students' choice of tests (and their results on those exams). We focus on allocation rules which are robustly implementable and stable for all specifications of preferences. $p(t, \{u_i, v_j\})$ denotes the allocation rule at a given type profile $t$ and preferences $\{u_i, v_j\}$. $p$ is \emph{ex-post stable} if, for all preferences $\{u_i, v_j\}$ and type vectors $t$, the match induced by $p(t, \{u_i, v_j\})$ is stable in the standard \cite{gale_1962_college} sense. 

We emphasize here that we have purposefully omitted student interim reporting constraints, in the sense that we have supposed preferences are completely pinned down by student ability. However, we can extend our model to the case where students jointly choose a Blackwell experiment about their type $\tau_i \in \Delta(\Delta(\mcal T_i))$ and report their preference function $u_i(a(i), t_i)$ (which may be their private information before reports). When this is the case, the deferred-acceptance algorithm we define below will both robustly incentivize information acquisition and be strategy-proof for students' preference reports (following \cite{roth_1982_the} and \cite{roth_1989_twosided}). For simplicity of exposition, we omit the second part from our notation and analysis, though we discuss how to accommodate it at the end of the proof of Proposition \ref{p: stable matching} in \blue{\ref{proof of p: stable matching}}\quad In the main body, we instead show that sender-proposing deferred acceptance can arise as the (essentially) unique robust, stable algorithm, even if we completely ignore the standard preference strategy-proofness desiderata. 

One immediate consequence of our grim-trigger robustness principle (Theorem \ref{t: grim trigger robustness principle}) is that robust allocation rules are implemented exactly by mechanisms which do not condition the resulting match on students' choice of experiment. 
This provides one potential justification for the credibility of the statement by college officials that they do not care about which admissions test is taken (e.g. SAT or ACT), just an applicant's score\footnote{Of course, we have abstracted from several other real-world considerations that administrators may be concerned about when deciding a matching system: imperfect availability of exams, persistent student private information about their ability, locational constraints, and so on. We have purposefully shut down these other channels to obtain a stylized model where we can parsimoniously understand the effect that matching algorithms have on student \emph{incentives} to take informative exams. Our results imply that, even if this was the only consideration a school administrator faced, student-proposing deferred acceptance is uniquely able to simultaneously handle equity and truthful reporting concerns.}. 
Among all allocation rules, a particular class of interest are those generated by some deferred acceptance algorithm. 

\begin{definition}
    The \emph{Sender-proposing deferred acceptance} allocation rule is the match \\$p(t, \{u_i, v_j\})$ resulting from the following algorithm. 
\begin{enumerate}[label=(\arabic*), ref={Sender-proposing deferred acceptance}]
    \item\label{deferred acceptance} Fix $t$. Suppose Sender $i$ has preferences over $\mcal J$ represented by $u_i(j, t_i)$, and Designer $j$ has strict preferences over $\mcal I$ represented by $v_j(i, t_i)$. 
    \item With these reporting preferences, run the Gale-Shapley deferred acceptance algorithm where students (senders) propose to schools. 
    Set the resulting match to be the outcome of $p^{SPDA}(t, \{u_i, v_j\})$.
    \item Repeat steps (1) and (2) for all possible $(t, \{u_i, v_j\})$. 
\end{enumerate}
\end{definition}

Step (3) need only be run for each $t$ and possible ordinal preference over $\{\mcal I, \mcal J\}$, since deferred acceptance is independent of changes in cardinal payoffs; in particular, this implies the algorithm terminates after finitely many steps. Designer-proposing deferred acceptance is defined similarly.

Our main result will be that the algorithm which implements the standard student-proposing deferred acceptance algorithm when all types are known (and otherwise allows grim-trigger punishments)\footnote{We show in the proof that it is not necessary for the mechanism to adopt such a drastic policy off-path, and in fact it is sufficient to give ``unilateral deviators'' the lowest priority in the student-proposing deferred acceptance algorithm. Intuitively, many punishments can work to discipline behavior since incentive constraints need not always bind.} is the only (up to allocation equivalence) algorithm which is stable and robustly implementable for all possible preferences. 

\begin{proposition}
    \label{p: stable matching}
   $p^{SPDA}$ is ex-post stable and robustly implementable. For any ex-post stable allocation rule $p$ satisfying $p \neq p^{SPDA}$ at any type vector for which more than one stable match exists, there exist preferences $\{u_i, v_j\}$ such that $p$ is not robust. 
\end{proposition}

The proof can be found in \blue{\ref{proof of p: stable matching}}. 
The key idea behind robustness is to use an alternative characterization of deferred acceptance, due to \cite{yannai}, to characterize a type-independent payoff lower bound for each sender, and then show the grim-trigger payoff is worse at every belief. Stability then follows by construction of the algorithm. 
We remark here that we can only give a partial converse: We require the (somewhat strong) assumption that the ex-post stable allocation rule differs from the SPDA allocation whenever possible. This hypothesis is satisfied by both Designer-proposing deferred acceptance and all ``compensation chain'' mechanisms (\cite{dworczak}) which find stable matches, and hence none of those algorithms are robust and stable for all preferences. However, we cannot rule out allocation rules where $p = p^{SPDA}$ sometimes but may also output some other stable match other times (though we are not aware of any algorithms which satisfy this condition). Thus, we see the converse as a reason to prefer student-proposing deferred acceptance among all common algorithms from a purely informational perspective (and in particular, a reason to prefer it over school-proposing deferred acceptance). 
Moreover, as the proof of Proposition \ref{p: stable matching} makes clear, we can generalize the converse to any algorithm $p$ which outputs a different outcome than SPDA for two players at least on two type profiles. This condition is needed since matchings rely on ordinal preferences but robustness relies on cardinal preferences. Without this condition, it is difficulty to get a uniform ordinal bound, in which case arbitrarily strong intensity of preferences becomes an issue.

\section{Discussion}
\makeatletter\def\@currentlabel{Section 6}\makeatother
\label{section_6}
In this paper, we introduced a model of contracting for information acquisition. We proved a revelation-like principle which highlighted the incentive costs of implementation in our model. We then turned to the question of prior-free implementation, and established an equivalence between robust implementation and experiment neutrality. Along the way, we develop some general technical machinery to characterize the incentive constraints which underlie (robust) implementation, which we hope is useful to future work that studies contracting for information acquisition. 

We next applied our results to two applications. First, we studied efficient good allocation. In contrast to strong impossibility theorems (even when the mechanism can use transfers to further screen agents) in settings where senders' valuations are their known private information, we give a possibility theorem for robust, efficient implementation in the ex-ante setting. We interpret our result as highlighting that the key friction in impossibility-like results is in information transmission and not acquisition. Moreover, our indirect implementation aligns sender reporting incentives in a way that is reminiscent of the Vickrey-Clarke-Groves mechanism, though without transfers, highlighting a natural way to robustly incentivize efficient information acquisition. 
Second, we considered a school choice setting with ex-ante certainty about ability, and characterize sender-proposing deferred acceptance in terms of its robust informational content. We interpret our result as implying that a school district (which faces heterogeneous preferences, many students, and rich preferences) which hopes to robustly implement a stable allocation should simply commit to student-proposing deferred acceptance. This will incentivize students to take maximally informative exams. 

There are several promising directions for future research. First, our implementation results extend to the case where the set of experiments is simply a lattice, but we are not sure how to characterize robust implementation in this setting, since a belief-based approach may not apply. Second, allowing Sender to have some imperfect, private information about their type may weaken Designer's commitment power but allow us to shed more light on the role of prior uncertainty in persuasion. Third, the setting where Sender's information acquisition is costly is natural. While we conjecture our results extend to the case where Sender costs are uniformly posterior separable and arbitrarily small, we do not know how to handle the case with fixed, exogenous costs.

\bibliography{cites.bib}

\appendix
\section*{Appendix A: Omitted Proofs}
\makeatletter\def\@currentlabel{Appendix A}\makeatother
\label{Appendix A}
\setcounter{lemma}{0}
\renewcommand{\thelemma}{A.\arabic{lemma}}

\setcounter{proposition}{0}
\renewcommand{\theproposition}{A.\arabic{proposition}}

\setcounter{definition}{0}
\renewcommand{\thedefinition}{A.\arabic{definition}}

\setcounter{corollary}{0}
\renewcommand{\thecorollary}{A.\arabic{corollary}}

\subsection*{A.I: Omitted Proofs in Section 3}
\makeatletter\def\@currentlabel{Appendix A.I.}\makeatother
\label{Appendix A.I}
\subsubsection*{PROOF OF THEOREM \ref{t: implementability}}
\label{proof of t: implementability}
\begin{proof}
    Clearly (2) implies (1). To show that (1) implies (2), suppose $p$ is induced by $(m, \tau^*)$ with $\tau^* \in \Sigma(m, \mu^0)$. If $\tau^*$ is fully-revealing, then we are done. Otherwise, consider the modified mechanism 
   \[ \hat m(\tau, \mu) =  \begin{cases} p(t) & \text{   if   } \tau = \hat \tau \text{  and  }\mu = \delta_t \in \supp(\hat \tau) \\ r(\mu) = \argmin_{\mcal A} \int u(a, t) d\mu(t) & \text{  else  } \end{cases} \]
   where $\hat \tau$ is the fully revealing experiment and $r(\mu)$ is the grim trigger correspondence. 
    Note $(\hat m, \hat \tau)$ induce $p(t)$ by construction, since for every state $t$, 
   \[\int \hat{m}(\hat{\tau}, \mu) d\tau(\mu|t) = \hat{m}(\hat{\tau}, \delta_t) d\tau(\delta_t|t) = \hat{m}(\hat{\tau}, \delta_t) = p(t)\]
   Moreover, for any $\tilde \tau$ which is not fully revealing, 
   \begin{align*}
       \iint u(\hat m(\tilde \tau, \mu), t) d\mu d \tilde \tau = \iint u(r(\mu), t) d\mu d\tilde \tau \leq \iint u(m(\tilde \tau, \mu), t) d\mu d \tilde \tau 
       \\ \leq \iint u(m(\tau^*, \mu), t) d\mu d \tau^* = \iint u(p(t), t) d\mu^0 = \iint u(\hat m(\hat \tau, \mu), t) d\mu d\hat \tau 
   \end{align*}
   The first inequality follows from the fact $m(\tilde \tau, \mu)$ is a feasible random action at $\mu$, and the second from the fact $\tau^* \in \Sigma(m, \mu^0)$. The penultimate equality follows since $(m, \tau^*)$ induce $p$. Thus, $\tilde \tau$ is not a profitable deviation for Sender, so $\hat \tau \in \Sigma(\hat m, \mu^0)$. 

Now we show (2) implies (3). Let $p$ be implemented by a direct mechanism $\hat m$ and fully revealing strategy $\tau^*$.
Thus, the deviation $\tau^0 = \delta_{\mu^0}$, which completely pools beliefs, must be unprofitable. Thus,
\[ \int u(p(t), t) d\mu^0 \geq \iint u(\hat m(\tau^0, \mu), t) d\mu d\tau^0 \geq \int u(r(\mu^0), t) d\mu^0 \]
where the first inequality uses the fact $p$ is induced by the direct mechanism $\hat m$, and the second from the fact $\hat m(\tau^0, \mu)$ was a feasible grim trigger punishment at the beliefs induced by $\tau^0$. 
Finally, (3) implies (1). Suppose the inequality holds, and let $\tilde \tau$ be any strategy where Sender deviates. Then, 
\begin{align*}
    \int u(p(t), t) d\mu^0 \geq \int u(r(\mu^0), t) d\mu^0 
    = \iint u(r(\mu^0), t) d\mu d\tau^0 \geq \iint u(r(\mu), t) d\mu d\tilde \tau
\end{align*}
The first inequality follows by assumption. The equality follows from Bayes plausibility. The second inequality follows from Blackwell's information theorem: $r(\mu^0)$ minimizes Sender $i$'s utility given posterior belief $\mu^0$, (alternatively, maximizes $-u(a, t)$) and $\tau^0$ is Blackwell less-informative than any other deviating experiment $\tilde \tau$. Thus, Sender does worse under $\tilde \tau$ than $\tau^0$, so $p$ is implementable. 
\end{proof}

\subsection*{A.II: Omitted Proofs in Section 4}
\makeatletter\def\@currentlabel{Appendix A.II.}\makeatother
\label{Appendix A.II}
\subsubsection*{PROOF OF THEOREM \ref{t: grim trigger robustness principle}}
\label{proof of t: grim trigger robustness principle}
\begin{proof}
We show that (1) and (2) are both equivalent to the pointwise incentive constraints in (3). 
First, (1) and (3) are equivalent. Clearly (1) implies (3) by Theorem \ref{t: implementability}, since robust implementation implies implementation at every full-support prior. That this extends to every belief $\mu \in \Delta(\mcal T)$ follows by taking limits and upper semicontinuity of the incentive constraints in beliefs (noting $r(\mu)$ is upper hemicontinuous). That (3) implies (1) follows similarly; take the shoot-the-sender mechanism from Theorem \ref{t: implementability} and note that it robustly implements the allocation regardless of the prior, so long as (3) is satisfied for all full-support priors. 

Next, (2) and (3) are equivalent. Both directions are done by contraposition. 
Suppose (3) fails for some belief $\tilde \mu$. Then 
\[ \int u(p(t), t) d \tilde \mu < \int u(r(\tilde \mu), t) d \tilde \mu\]
Now fix any direct, belief-based grim trigger mechanism $m$. For $m$ to induce $p$, it must be that $m(\delta_t) = p(t)$ for all states $t$. 
Fix any full-support prior belief $\bar \mu$ and let $\tau^*$ be the fully revealing experiment which is Bayes-plausible at $\bar \mu$. 
Construct the following unilateral deviation $\tau'$: split $\bar \mu$ into the belief $\tilde \mu$ which fails condition (3) and some complementary belief $\tilde \mu^c$ chosen to satisfy $\bar \mu = \beta \tilde \mu + (1 - \beta)\tilde \mu^c$ (i.e. to maintain Bayes plausibility at $\bar \mu$). 
Since the prior is full-support, such weights $\beta > 0$ can always be found for some $\tilde \mu^c$. Now suppose $\tau'$ induces belief $\tilde \mu$ with probability $\beta$ and with complementary probability fully reveals the state. By choice of $\beta$, this satisfies Bayes plausibility at $\bar \mu$ and thus this is a feasible strategy for player $i$. 
But then 
\begin{align*}
 \mcal U(\tau', m) - \mcal U(\tau^*, m) =  \beta\left( \int u(r(\tilde \mu), t) d\tilde \mu - \int u(p(t), t) d\tilde \mu \right) > 0
\end{align*}
since $\tau_i'(\tilde \mu_i) = \beta > 0$ and the supposition implies the term inside the parentheses is positive. This implies $\tau'$ is a profitable deviation.

The converse. Assume $p$ is not implementable by a direct, belief-based grim trigger mechanism $m$ but that Condition (3) holds. Then there exists prior $\mu^0$ and strategy $\tilde \tau$ (Bayes plausible at $\mu^0$) that is strictly more profitable than $\tau^*$. Let $\mcal B$ be the set of nondegenerate beliefs; note $\tilde \tau(\mcal B) > 0$ by assumption. Then we have 
\begin{align*}  
\mcal U(\tilde \tau, m) = \iint_{\mcal B} u(r(\mu), t) d\mu d\tilde \tau + \iint_{\mcal B^c} u_i(p(t), t) d\mu d\tilde \tau \\
\leq \iint_{\mcal B} u(p(t), t) d\mu d \tilde \tau + \iint_{\mcal B^c} u(p(t), t) d\mu d\tilde \tau = \iint u(p(t), t) d \mu d \tilde \tau = \mcal U(\tau^*, m)
\end{align*}
where the inequality holds by Condition (3) and the second equality by Bayes plausibility. This implies that $\tilde \tau$ is not a profitable deviation, a contradiction. This finishes the proof. 
\end{proof}

\subsubsection*{PROOF OF PROPOSITION \ref{p: indifference beliefs}}
\label{proof of p: indifference beliefs}
\begin{proof}
First, note the grim-trigger correspondence $r: \Delta(\mcal T) \rightrightarrows \mcal A$ is nonempty, compact-valued, and varies upper hemi-continuously by Berge's theorem. 
We next formalize the geometric characterization of $K(a)$. 

\begin{lemma}
\label{kir_convex}
    $K(a)$ is convex and compact for each $a$, and $\U_a K(a) = \Delta(\mcal T)$. 
\end{lemma}
\begin{proof}
    Clearly, $\{K(a)\}$ covers $\Delta(\mcal T)$ since $r(\mu)$ is always nonempty-valued. Convexity follows from the sure-thing principle; for any $\mu', \mu'' \in K(a)$, we have 
   \begin{align*}
       \int u(a, t) d\mu \leq \int u(a' ,t) d\mu \text{  for all } a' \in \mcal A, \mu \in \{\mu', \mu''\} 
       \\ \implies \int u(a, t) d \left(\beta \mu' + (1 - \beta)\mu''\right) \leq \int u(a', t) d \left(\beta \mu' + (1 - \beta)\mu''\right)
   \end{align*}
    for all $a' \in \mcal A$ and $\beta \in [0, 1]$ by linearity of expectation.
    Finally, compactness. It is sufficient for $K(a)$ to be closed since $\Delta(\mcal T)$ is compact.
    Fix $\{\mu^n\} \subset K(a)$ where $\mu^n \to \mu$. Then $a \in r(\mu^n)$ for all $n$, so then $a \in r(\mu)$ by upper hemicontinuity of the grim-trigger correspondence. 
\end{proof}

Lemma \ref{kir_convex} implies the convex program on the left hand side of the proposition can be rewritten as the ``two-stage'' optimization problem 
\begin{align*}
\min_{\mu \in \Delta(\mcal T)} \left\{ \int u(p(t), t) d\mu - \int u(r(\mu), t) d\mu \right\} = \min_{a \in \mcal A} \left\{ \min_{\mu \in K(a)}  \left\{ \int u(p(t), t) d\mu - \int u(r(\mu), t) d\mu \right\} \right\}.
\end{align*}
But now the inner program is linear in the objective $\mu$, and is minimized over a compact and convex set (by Lemma \ref{kir_convex}); Bauer's theorem of the maximum then implies it must include an extreme point of $K(a)$. 

Next, we want to show $\mcal E$ is finite. Theorem 19.1 of \cite{convex_rockafellar} implies it is sufficient for each $K(a)$ to be the intersection of finitely many half-spaces (interpreting $K(a) \subset \bb{R}^{|\mcal T|})$. Note 
\[ K(a)= \Delta(\mcal T) \cap \left( \A_{a' \in \mcal A \setminus \{a\}} \left\{ \mu : \int u(a', t) - u(a, t) d \mu \geq 0 \right\} \right) \]
since for each $a'$, the set of beliefs (e.g. vectors) which satisfy the inequality is a half space in $\bb{R}^{|\mcal T|}$. 
Moreover, $\Delta(\mcal T)$ has finitely many extreme points and is thus the intersection of finitely many halfspaces. Together, this implies $K(a)$ is the finite intersection of half-spaces. 

Finally, we want to show 
\[ \U_{a \in \mcal A} \ext(K(a)) \subset \ext(\Delta(\mcal T)) \cup \mcal D \]
Fix some $a \in \mcal A$ and belief $\tilde \mu \in \text{ext}K(a) \setminus \text{ext}(\Delta(\mcal T))$. Since $\tilde \mu$ is not an extreme point of $\Delta(\mcal T)$, it has nondegenerate support. If $\tilde \mu \not\in \mcal D$, then for every $a' \neq a$, $\tilde \mu \not\in K(a')$. Since each $K(a)$ is closed, there exists an open set $U_{a'} \subset \Delta(\mcal T)$ such that $\tilde \mu \in U_{a'}$ but $U_{a'} \cap K(a') = \varnothing$. Then $U = \A_{a' \neq a} U_{a'}$ is open (it is a finite intersection of open sets) containing $\tilde \mu$ and disjoint from $\U_{a' \neq a} K(a')$. Because $\tilde \mu$ is an extreme point of $K(a)$ not on the boundary of $\Delta(\mcal T)$, $K(a)^c \cap U$ contains some belief $\mu' \in \Delta(\mcal T)$. But then $\mu' \not\in \U_{a} K(a)$, contradicting the fact the sets $K(a)$ cover $\Delta(\mcal T)$. 
\end{proof}

\subsection*{A.III: Omitted Proofs in Section 5}
\makeatletter\def\@currentlabel{Appendix A.III.}\makeatother
\label{Appendix A.III}

\subsubsection*{PROOF OF PROPOSITION \ref{p: homogeneous least favorites}}
\label{proof of p: homogenous least favorites}
\begin{proof}
Suppose Sender preferences satisfy homogeneous least favorites and $a_{LF}$ is the constant selection in $r(\cdot)$. Then 
\[ \int u(p(t), t) - u(r(\mu), t) d\mu = \int u(p(t), t) - u(a_{LF}, t) d\mu \geq 0\]
for all $\mu$, so any $p$ is robustly implementable by Theorem \ref{t: grim trigger robustness principle}. 

The converse. Suppose Sender preferences do not satisfy homogenous least favorites. Define $p(t) = \argmin_{a} u(a, t)$ to be Sender's least favorite action. Fix any full-support belief $\mu$, and not that by the failure of homogenous least favorites we have that for all $a$,
\[ \int u(a, t) d\mu > \int u(p(t), t) d\mu \implies \int u(r(\mu), t) d\mu > \int u(p(t), t) d\mu\]
since there are finitely many actions. Thus pointwise IC fails at $\mu$. 
\end{proof}

\subsubsection*{PROOF OF PROPOSITION \ref{p: monotone implementability}}
\label{proof of p: monotone implementability}
\begin{proof}
The result is clear if homogenous least favorites is satisfied. Assume not, so $\mcal D$ is nonempty. Fix indifference belief $\mu \in \mcal D$ such that $r(\mu) = \{a_1, a_2\}$. Suppose we select $a_2$ from the correspondence without loss of generality, by symmetry. Some algebra implies we can write pointwise incentive constraints as 
\begin{align*}
    \int u(p(t), t) - u(r(\mu), t) d\mu = \int u(p(t), t) - u(a_2, t) d\mu 
    \\ = \int (u(a_1, t) - u(a_2, t)) A(t) d\mu \text{  where  } A(t) = \bb{P}(p(t) = a_1)
\end{align*}
is the probability of receiving outcome $a_1$ at state $t$. The first equality uses the fact we set $r(\mu) = a_2$, and the second from the fact there are only two outcomes, so $u(p(t), t) = u(r(\mu), t)$ whenever $p(t) = a_2$ and the difference zeros. 

For any $u(a, t)$, we can find an order $\succsim_{\mcal T}$ on types such that when actions are ordered so that $a_1 \succsim a_2$, $u(a_1, t) - u(a_2, t)$ is single-crossing. While not necessary for the proof, the following observation is without loss of generality and will simplify some computations below. 

In particular, the above observation ensures there is some minimal element $\bar t$ such that $t \succsim^{\mcal T} \bar t$ if and only if $u(a_1, t) \geq u(a_2, t)$. Thus, we can rewrite the payoff differential as 
\[ \int_{t \succsim^{\mcal T} \bar t} A(t)\left[ u(a_1, t) - u(a_2, t) \right] d\mu + \int_{\bar t \succ^{\mcal T} t} A(t)\left[ u(a_1, t) - u(a_2, t) \right]. \]
We have chosen $\bar t$ so that the first integrand is positive and the second is negative. Since homogenous least favorites is not satisfied, both sets we are integrating over are not empty. We can bound the expression above by 
\begin{align*}
    A(t_{\max}) \int_{t \succsim^{\mcal T} \bar t} \left[ u(a_1, t) - u(a_2, t) \right] d\mu + A(t_{\min}) \int_{\bar t \succ t} \left[ u(a_1, t) - u(a_2, t) \right] d\mu 
    \\ \geq A(t_{\min}) \int  \left[ u(a_1, t) - u(a_2, t) \right] d\mu = 0 
\end{align*}
where $t_{\max}$ and $t_{\min}$ solve the infimum and supremum condition in the theorem statement, respectively. The inequality follows by noting (1) $t_{\max} \succsim^{\mcal T} t_{\min}$ and (2) $A(t_{\max}) \geq A(t_{\min})$ by assumption, while the final equality uses the fact $\mu$ was an indifference belief.

The converse. Suppose the condition fails, so that there exists $\bar t$ and $\und t$ such that $A(\bar t) < A(\und t)$ but $u(a_1, \bar t) \geq u(a_2, \bar t)$ and $u(a_1, \und t) < u(a_2, \und t)$. The computation above then implies at the (unique) indifference belief $\mu \in \mcal D$ which supports only $\{\bar t, \und t\}$,   
\begin{align*}
    A(\bar t)\left[ u(a_1, \bar t) - u(a_2, \bar t) \right] \mu(\bar t) + A(\und t)\left[ u(a_1, \und t) - u(a_2, \und t) \right] \mu(\und t)
    \\ \leq A(\und t) \left( \int \left[ u(a_1, t) - u(a_2, t) \right] d\mu \right) = 0.
\end{align*}
Thus, pointwise incentive constraints fail at $\mu$, so $p$ is not robustly implementable. 
\end{proof}

\subsection*{A.IV: Omitted Proofs in Section 6}
\makeatletter\def\@currentlabel{Appendix A.IV.}\makeatother
\label{Appendix A.IV}

\paragraph{PROOF OF PROPOSITION \ref{p: efficient allocation}}
\label{proof of p: efficient allocation}
    \begin{proof}
    Fix any prior $\mu^0$ and any Sender $i$. We show pointwise incentive constraints hold. Since the prior was arbitrary, this implies full robustness by Theorem \ref{t: grim trigger robustness principle}. 
We can rewrite agent $i$'s pointwise incentive constraints as 
\begin{align*}
    &\iint u_i(p^*(t), t_i) - u_i(r_i(\mu_i), t_i) d\mu_{-i}^0 d\mu_i   \\
    =  & \iint [ u_i(p^*(t), t_i) - u_i(p_{-i}(t_{-i}), t_i)] \mathbf{1} \{p^*(t) = i\} d\mu_{-i}^0 d\mu_i  \\ 
    & + \iint [ u_i(p^*(t), t_i) - u_i(p_{-i}(t_{-i}), t_i)] \mathbf{1} \{p^*(t) \neq i\} d\mu_{-i}^0   d\mu_i \\
    & + \iint u_i(p_{-i}(t_{-i}), t_i) - u_i(r_i(\mu_i), t_i) d\mu_{-i}^0 d\mu_i. 
\end{align*}
Recall $p_{-i}(t_{-i})$ is the allocation rule assigned by the leave-one-out mechanisms on deviations. 

We treat each term in the rewritten sum separately. The first term conditions on the event $p^*(t) = i$. Since preferences satisfy weakly negative externalities, this must imply that $u_i(p^*(t), t_i) \geq 0$, as otherwise $\sum_{j \in I} u_j(i, t_j) < 0$ and the efficient allocation rule would have the good thrown away at $t$.
But then because $u_i(p_{-i}(t_{-i}), t_i) \leq 0$ for all $p_{-i}(t_i)$ again by our weakly negative externalities assumption, it must be that    
\[ \iint [ u_i(i, t_i) - u_i(p_{-i}(t_{-i}), t_i)] \mathbf{1} \{p^*(t) = i\} d\mu_{-i}^0 d\mu_i \geq 0 \] 
as a nonnegative term less a nonpositive term is nonnegative.

The second term conditions on the event $p^*(t) \neq i$. We show 
\[ u_i(p^*((t_{-i},t_i)),t_i) \geq u_i(p_{-i}(t_{-i}), t_i).\]
Suppose this pointwise inequality was false, so that $u_i(p^*(t),t_i) < u_i(p_{-i}(t_{-i}), t_i)$. This implies $p^*(t) \neq p_{-i}(t_{-i})$. 
    Let $j = p^*(t), k = p_{-i}(t_{-i})$. As $j$ maximizes total surplus, 
    \[\sum_{\ell \in \mcal I} u_\ell(j, t_\ell) \geq \sum_{\ell \in \mcal I}  u_\ell(k, t_\ell).\]
    Yet we also know that $u_i(j, t_i) < u_i(k, t_i)$. Combining these inequalities implies
    \[\sum_{\ell \in \mcal I} u_\ell(j, t_\ell) - u_i(j, t_i) > \sum_{\ell \in \mcal I} u_\ell(k, t_\ell) - u_i(k, t_i).\]
    which implies
    \[\sum_{\ell \in \mcal I \setminus \{i\}} u_\ell(j, t_\ell) > \sum_{\ell \in \mcal I \setminus \{i\}} u_\ell(k, t_\ell).\]
    However, $j$ yields strictly higher total surplus among senders other than $i$, contradicting that $k = p_{-i}(t_{-i})$ maximizes the leave-one-out aggregate.
    Integrating this pointwise inequality then implies 
    \[ \iint [u_i(p^*(t), t_i) - u_i(p_{-i}(t_{-i}), t_i)] \mathbf{1}\left\{p^*(t) \neq i \right\} d\mu_{-i}^0(t_{-i}) d\mu_i \geq 0 \]
    Finally, note that $p_{-i}(t_{-i})$ is always a feasible grim-trigger punishment, and hence the last term must always be nonnegative. Adding together these three nonnegative terms implies $p^*$ always satisfies pointwise incentive constraints, showing $p^*(t)$ is robust. Moreover, by noting that the deviation loss under the leave-one-out mechanism is exactly the first two terms, and adapting a version of the pointwise incentive compatibility argument of Theorem \ref{t: grim trigger robustness principle}, it should be clear the leave-one-out mechanism robustly implements $p$. 
\end{proof}

\paragraph{PROOF OF PROPOSITION \ref{p: stable matching}}
\label{proof of p: stable matching}
\begin{proof}
For the summary of incentive constraints with multiple senders (which we use throughout here), see \blue{\ref{Appendix B}. }
Fix senders $(\mcal I, \mcal J)$, type spaces $\mcal T$ and preferences $\{u_i, v_j\}$, and prior $\mu^0$. Stability of $p^{SPDA}$ follows from \cite{roth_1989_twosided} and the definition of $p^{SPDA}$. 
For robustness, fix sender $i$ and recall sender $i$'s match is given by the following algorithm (see \cite{yannai}). 
    \begin{enumerate}
		\item Run Designer-proposing deferred acceptance on the other $|\mcal I| - 1$ senders and $n$ Designers at type $t$ and preferences $\{u_i, v_j\}$.
		\item Allow individual $i$ to pick from any unmatched Designer or Designer that would prefer $i$ over their current match under Step (1). 
    \end{enumerate}
Denote the induced ``leftover allocation rule'' from Step (1) by $\hat p_i(t_{-i}, \{u_i, v_j\})$, where we leave Sender $i$ out. As $\hat p_i(t_{-i}, \{u_i, v_j\})$ is always feasible for option $i$, we have 
\[ u_i(p^{SPDA}(t_i, t_{-i}, \{u_i, v_j\}), t_i) \geq u_i(\hat p_i(t_{-i}, \{u_i, v_j\}), t_i) \]
for all type profiles $(t_i, t_{-i})$. Since $\hat p_i(t_{-i}, \{u_i, v_j\})$ is independent of $t_i$, it is a feasible punishment, so $r_i(\mu_i)$ is (weakly) worse for $i$ than $\hat p_i(t_{-i}, \{u_i, v_j\})$. 
Putting these together and integrating against $\mu_{-i}^0$ and $\mu_i$ (suppressing dependency on $\{u_i, v_j\}$ gives 
\begin{equation*}
    \begin{split}
        \iint u_i(p^{SPDA}(t_i, t_{-i}), t_i) d\mu_{-i}^0 d\mu_i \geq \iint u_i(\hat p_i(t_{-i}), t_i) d\mu_{-i}^0 d\mu_i \geq \iint u_i(r_i(\mu_i), t_i) d\mu_i
    \end{split}
\end{equation*}
so $p^{SPDA}$ is interim robust at $\mu^0$. As $\mu^0$ was arbitrary, Lemma \ref{full partial equivalence} implies $p^{SPDA}$ is robust. 

Note we can also accommodate the model where agents jointly report type-dependent preference functions $u_i$ as well. Since $\hat p_i)$ does not depend on $i$'s reported preferences, jointly deviating by choosing a non-truthful $\tau$ and preference $\hat u_i \neq u_i$ cannot increase their utility.\footnote{This also implies strategy-proofness for preference reporting in our setting.} 

The converse. Fix any type space $\mcal T$, and players $\mcal I$ where $|\mcal I| \geq 2$ and $|\mcal T_i| \geq 2$ for each $i \in \mcal I$, and full-support prior $\mu^0$. Suppose $p(t, \{u_i, v_j\}) \neq p^{SPDA}(t, \{u_i, v_j\})$  but is always stable. 
We construct preferences $\{u_i, v_j\}$ such that (1) there are (only) two stable matches, attained by $p^{SPDA}$ and $p^{RPDA}$ (Designer proposing DA) such that for some sender $p^{RPDA}$ is not robustly implementable at $\mu^0$. For any $i, j > 2$, suppose Sender and Designer preferences are: 
\vspace{-0.6em}\begin{table}[h!]\centering
    \begin{tabular}{|c|c|c|}
      \hline &  $r_j, j = i$  & $r_j, j \neq i$ \\ \hline
      $t_i \in \mcal T_i$ & 100, 100 & $\varepsilon_j$,0  \\ \hline
    \end{tabular}
\end{table}

where $\varepsilon_j > 0$ are arbitrarily small tiebreaking indices to ensure student preferences are strict. 

Now suppose instead that $t^1, t^2 \in \mcal T_1, \mcal T_2$. Let senders $1,2$ and Designers $r_1,r_2$ have identity-independent payoffs $u_i(j, t_i), v_j(i, t_i)$ as described in the following matrix:
\begin{table}[h]\centering
    \begin{tabular}{|c|c|c|c|}
      \hline & $r_1$ & $r_2$ & $r_j, j \neq 1, 2$  \\ \hline
      $t^1$ & $1,10$ & $5,5$ & $10, 11$ \\ \hline
      $t^2$ & $5,5$ & $1,10$ & $10, 11$\\ \hline
      $t\neq t^1, t^2$ & $-100, -100$ & $-100, 10$ & $-100, -100$\\ \hline
    \end{tabular}
\end{table}

This implies Sender $i$ is matched with Designer $i$ whenever $i > 2$. We can now focus on the sub-matching problem with only senders $\{1, 2\}$ and Designers $\{r_1, r_2\}$. 
The unique indifference belief for these senders is $(\frac12, \frac12)$, and they are indifferent between $r_1$ and $r_2$.

Suppose senders $i = 1, 2$ had prior beliefs at their indifference belief. Then 
\begin{itemize}
    \item With probability $1/2$, the senders have the same type and have total utility $1+5=6$ regardless of how they are matched;
    \item With probability $1/2$, there is one Sender of type $t^1$ and one Sender of type $t^2$. Under $p^{SPDA}$, Sender of type $t^1$ is matched with $r_2$ while Sender of type $t^2$ is matched with $r_1$. Under $p^{RPDA}$, Sender of type $t^1$ is matched with $r_1$ while Sender of type $t^2$ is matched with $r_2$. As this is the only other stable match, $p = p^{RPDA}$ in this (sub-) market.
\end{itemize}
Thus, total expected utility generated from these two senders is $\frac{1}{2}(6 + 2) = 4$ and at least one Sender receives expected utility less than $3$. However, the grim-trigger punishment at either Sender's indifference belief yields an utility of $\frac{1}{2}(1 + 5) = 3$ so pointwise incentive constraints fail for at least one Sender at the prior $\mu_1^0 = \mu_2^0 = (\frac{1}{2}, \frac{1}{2})$. Thus, $p$ is not robust.
\end{proof}

\section*{Appendix B: Many Senders}
\makeatletter\def\@currentlabel{Appendix B}\makeatother
\label{Appendix B}

\setcounter{lemma}{0}
\renewcommand{\thelemma}{B.\arabic{lemma}}

\setcounter{proposition}{0}
\renewcommand{\theproposition}{B.\arabic{proposition}}

\setcounter{definition}{0}
\renewcommand{\thedefinition}{B.\arabic{definition}}

\setcounter{corollary}{0}
\renewcommand{\thecorollary}{B.\arabic{corollary}}

\subsection*{B.I: Interim Payoffs and Interim Robustness}
\makeatletter\def\@currentlabel{Appendix B.I}\makeatother
\label{Appendix B.I}
Suppose there are $|\mcal I| < \infty$ senders. Throughout, we will subscript notation (e.g. $\tau_i, u_i, t_i$, etc.) to refer to individual-by-individual objects and will drop subscripts when referring to the entire vector of strategies, payoffs, types, and so on.
We assume throughout that types are drawn independent of each other according to their prior. We also assume preferences depend only on each Sender's state, so $u_i(a, t) = u_i(a, t_i)$\footnote{This is made for notational convenience only; our main results extend to the case with interdependent values, at the cost of introducing unnecessary notation. Since our applications satisfy the independent values assumption, we omit it from the appendix.}. 

A mechanism $m: \prod_i (\Delta(\Delta(\mcal T_i)) \times \Delta(\mcal T_i)) \to \Delta(A)$ now maps a vector of experiments and beliefs into a (random) action. 
Given a strategy $\tau$ and mechanism $m$, we define agent $i$'s \emph{interim payoff} as 
\[ \mcal U_i(\tau, m) = \iint u_i(a, t) dm(\tau, \mu) d\mu d\tau. \]
Mechanism $m$ implements a strategy $\tau$ at prior $\mu^0$ if, for all $i$, 
\[ \tau_i \in \argmax_{\tau_i' \in \Delta(\Delta(\mcal T_i)): \bb{E}_{\tau_i'}[\mu] = \mu_i^0} \left\{ \mcal U_i(\tau_i', \tau_{-i}, m) \right\}. \]
That is, $\tau$ is a Nash equilibrium of the complete-information game where each agent's strategy is the set of Bayes plausible distributions of posteriors and payoffs are given by $\mcal U_i(\tau, m)$ (for fixed $m$). As before, we let $\Sigma(m, \mu^0)$ be the set of strategies implemented by a mechanism $m$. 
Given $m$ and $\tau \in \Sigma(m, \mu^0)$, the induced allocation rule $p_{m, \tau}$ is defined in the same way as before; similarly, we say $m$ implements $p$ at $\mu^0$ if there exists $m, \tau \in \Sigma(m, \mu^0)$ such that $p_{m, \tau} = p$. Given these adapted definitions, it should be clear that Theorem \ref{t: implementability} applies to the setting with many players as well. 

Definition \ref{d: robust definition} (robustness) is in principle unchanged. However, because there are many agents, we may need to consider their beliefs about the prior beliefs of other agents, as this will vary the allocation rule and thus their ex-ante payoffs. It can be complicated to jointly consider possible deviations for some agents and changes in other agents' prior beliefs. Consequently, it will be helpful to define the following notion of ``interim'' robustness (for some Sender $i$ and some belief $\mu_{-i}^0$ of other senders), where we fix the prior beliefs of all other senders except Sender $i$. 

\begin{definition}
    A mechanism $m$ is \emph{interim robust} at belief $\mu^0$ if, there exists $p$ such that for all $i$ and beliefs $\mu_i \in \Delta(\mcal T_i)$, $m$ implements $p$ at all priors of the form $(\mu_i, \mu_{-i}^0)$. 
    An allocation rule $p$ is \emph{interim robust} at belief $\mu^0$ if, there is a mechanism $m$ implementing $p$ which is interim robust at $\mu^0$. 
\end{definition}

\begin{lemma}
    \label{full partial equivalence}
 $p$ is robust if and only if it is interim robust at every belief $\mu^0$. 
\end{lemma}
\begin{proof}
    Suppose $p$ is robust. Then there is a robust mechanism $m$ which implements it. Clearly, this implies there is an interim robust mechanism $m$ which implements it at $\mu^0$.
    Conversely, suppose $p$ is interim robust at every belief $\mu^0$. This implies there exists a sequence of mechanisms $m^{\mu^0}$ which interim robustly implement $p$ at belief $\mu^0$. Define the mechanism $m$ as $m(\tau) = m^{\mu^0}(\tau) \text{  if  } \bb{E}_\tau[\mu] = \mu^0$, noting this completely specifies a mechanism. Then for any prior $\mu^0$, $m$ implements $p$, and hence both $m$ and $p$ are robust. 
\end{proof}

It should be clear that with one Sender, interim robustness and robustness exactly coincide. With more than one Sender, we verify robustness by proving interim robustness for each prior $\mu^0$ and then appealing to Lemma \ref{full partial equivalence} (this is the approach we take in the applications in the applications). Given this result, it will be useful to define the interim utilities given by an allocation rule $p$ (resp. a direct mechanism that implements $p$) and prior $\mu^0$ as 
\[ U_i(p; t_i, \mu^0) = \int u_i(p(t_i, t_{-i}), t_i) d\mu_{-i}^0(t_{-i}). \]
Having cast the problem in the interim, we can now prove analogies of the main results in the text, which hold in the interim with the defined interim  payoffs at some prior. 

\begin{definition}
    A mechanism $m$ is a \emph{direct, belief-based, grim-trigger mechanism} if
    \begin{enumerate}
        \item For every $\mu^0$ there exists $\tau \in \Sigma(m, \mu^0)$ such that $\tau$ is truthful. 
        \item $m(\tau, \mu) = m(\tau', \mu)$ for all $\tau, \tau'$ and $\mu$. 
        \item For any $\mu$ where $\#\{i : \supp(\mu_i) > 1\} = 1$, $m(\mu_i, \delta_{t_{-i}}) = r_i(\mu_i) \in \argmin_{a \in \mcal A} \left\{ \int u_i(a, t_i) d\mu_i\right\}$. 
    \end{enumerate}
\end{definition}

\begin{proposition}[Interim Grim-Trigger Robustness Principle]
TFAE. 
\begin{enumerate}
    \item An allocation rule $p$ is interim robust at belief $\mu^0$. 
    \item $p$ is implementable by a direct, belief-based, grim-trigger mechanism at $\mu^0$. 
    \item For each $i \in \mcal I$ and $\mu_i \in \Delta(\mcal T_i)$, 
    \[ \int U_i(p; t_i, \mu^0) d\mu_i \geq \int u_i(r_i(\mu_i), t_i) d\mu_i. \]
\end{enumerate}
\end{proposition}

Analogous version of Proposition \ref{p: indifference beliefs} and Proposition \ref{p: homogeneous least favorites} hold as well with the interim utilities (and interim robustness) instead of payoffs for just one Sender. Note that this iteration with many senders places more restrictions on robustness: An allocation rule $p$ is robust only if the induced interim allocation rule $\bb{E}_{\mu_{-i}^0}[p(t)]$ is robust given induced interim utilities $U_i(p; t_i, \mu^0)$. The generalized version of Proposition \ref{p: homogeneous least favorites} thus places restrictions on the degree to which the allocation rule $p$ can vary Sender $i$'s payoff as a function of some other Sender $j$'s report. 

\subsection*{B.II: Two-Decomposability}
\makeatletter\def\@currentlabel{Appendix B.II}\makeatother
\label{Appendix B.II}

That the allocation rule cannot vary payoffs too substantially for Sender $i$ as Sender $j$'s type changes may naturally be restrictive in good allocation rule problems. In this section, we give an adapted version of Proposition \ref{p: monotone implementability}, which generalizes the binary action assumption. The key idea is that from each Sender's perspective, it is as-if they are facing a binary action problem.  

\begin{definition}
  Sender preferences $\{u_i\}$ are \emph{two-decomposable} on $\mcal A$ if, for each Sender $i$, there exist actions $a_1^i$ and $a_2^i$ such that for all $a \in \mcal A$,  
    \[ u_i(a, \cdot) \in \left\{ u_i(a_1^i, \cdot), u_i(a_2^i, \cdot) \right\}. \]
\end{definition}

We say preferences $\{u_i\}$ are \emph{regular} if (1) they are two-decomposable on $\mcal A$, and (2) there exist orders $(\succsim_i^{\mcal A}, \succsim_i^{\mcal T_i})$ such that $u_i(a, t_i)$ is single-crossing when we order $(\mcal A, \mcal T_i)$ by $(\succsim_i^{\mcal A}, \succsim_i^{\mcal T_i})$. Without loss of generality, suppose $a_1^i \succsim_i^{\mcal A} a_2^i$ for each $i$. Let $\mcal A_i^+ = \{a : u_i(a_1^i, t_i) = u_i(a, t_i) \text{  for all   } t_i\}$ and $\mcal A_i^- =\{a : u_i(a_2^i, t_i) = u_i(a, t_i) \text{  for all   } t_i\}$. If preferences are two-decomposable then $\mcal A_i^+ \cup \mcal A_i^- = \mcal A$ for all $i$. Given these definitions, we can suitably modify the proof of Proposition \ref{p: monotone implementability} to obtain the following result.

\begin{proposition}
\label{two decomposable theorem}
    Let $\{u_i\}$ be regular. Then $p$ is robust if and only if for all senders $i$,
\[ \inf_{t_i: u_i(a_1^i, t_i) \geq u_i(a_2^i, t_i)} \left\{ \int \bb{P}(p(t) \in \mcal A_i^+) d\mu_{-i}^0 \right\} \geq \sup_{t_i: u_i(a_1^i, t_i) < u_i(a_2^i, t_i)} \left\{ \int  \bb{P}(p(t) \in \mcal A_i^+)  d\mu_{-i}^0  \right\}. \]
\end{proposition}

This generalization has important implications in the good allocation setting (see, for example, Proposition \ref{p: efficient allocation}). In particular, if there are no allocation externalities, then preferences are two decomposable with $\mcal A_i^+ = \{i\}$ and $\mcal A_i^- = \{j : j \neq i\}$. Proposition \ref{two decomposable theorem} then implies that all (and only) interim ordinally monotone allocation rules are implementable. It is easy to show that if $u_i(i, t_i) - u_i(j, t_i)$ is single-crossing then $p^*(t)$ is interim monotone, providing a much simpler proof of efficient implementation in the setting without allocation externalities.

\section*{Appendix C: (Non)-Revelation Principles for Belief-Based Mechanisms}
\makeatletter\def\@currentlabel{Appendix C}\makeatother
\label{Appendix C}

\setcounter{lemma}{0}
\renewcommand{\thelemma}{C.\arabic{lemma}}

\setcounter{proposition}{0}
\renewcommand{\theproposition}{C.\arabic{proposition}}

\setcounter{definition}{0}
\renewcommand{\thedefinition}{C.\arabic{definition}}

\setcounter{corollary}{0}
\renewcommand{\thecorollary}{C.\arabic{corollary}}
\label{Appendix C}

We show the set of belief-based direct mechanisms does not admit a revelation principle: It does not trace out the set of all allocation rules implemented by belief-measurable mechanisms. 
This highlights that the equivalence between (1) and (2) in Theorem \ref{t: grim trigger robustness principle} relies crucially on the restriction to direct mechanisms. 
Fix a prior $\mu^0$. Let $\mcal P^R$ be the set of robust allocation rules, $\mcal P^B$ the set of allocation rules which can be implemented by a direct belief-based mechanism at $\mu^0$, and $\mcal P^E$ the set of all implementable allocation rules at $\mu^0$. 

For shorthand of notation, we will refer to an \emph{economic environment} as a tuple $\{\mcal T, \mcal A, u, \mu^0\}$ where $u: \mcal A \times \mcal T \to \mathbb{R}$ and $\mu^0 \in \Delta(\mcal T)^o$.
We show that $\mcal P^R \subsetneq \mcal P^B \subsetneq \mcal P^E$. First, 

\begin{proposition}
\label{behavioral_with_loss}
   There exists a tuple $\{\mcal T, \mcal A, u, \mu^0\}$ such that at $\mu^0$, $\mcal P^R \subsetneq \mcal P^B$. 
\end{proposition}
\begin{proof}
    Suppose $\mcal A = \{0, 1\}$, $\mcal T = \{0, 1\}$, and payoffs are given by $u(a_j, t_k) = \mathbf{1}_{j \neq k}$. 
    To help clarify ambiguity, we will refer to the action $a = i$ as $a_i$ and the state $t = i$ as $t_i$ (for $i \in \{0, 1\}$).
  Set $\mu = \mu^0(t_1)$ to be the prior probability the state is $t_1$. Let $\mu = \frac14$. Note the grim-trigger correspondence is given by $r(\mu) = a_0$ if $\mu < \frac12$, $a_1$ if $\mu > \frac12$, and $\Delta(\{a_0, a_1\})$ is $\mu = \frac12$. 
    Consider the mechanism $m(\mu)$ defined to be $r(\mu)$ except on $\{0, \frac12, 1\}$, and suppose $m(0) = \frac34 a_0 + \frac14 a_1$, $m(1) = \frac14 a_0 + \frac34 a_1$, and $m(\frac12) = a_1$.  
    Sender's indirect utility from $m$ and its concavification, are graphed below in Figure \ref{B1_fig1}. 

\begin{figure}[h]
\centering
    \begin{tikzpicture}[scale=.85]
    
    \draw[very thick] (0,0) -- (8,0);
    \draw[very thick, ->] (0,-0.1) -- (0,5.5);
    \draw[very thick, ->]  (8,-0.1) -- (8,5.5);
    \node[] at (0,-0.5) {$\mu = 0$};
    \node[] at (8,-0.5) {$\mu = 1$};
    \node[] at (-1,5) {$u(a,t_0)$};
    \node[] at (9,5) {$u(a,t_1)$};

    \draw[very thick] (0,0) -- (8,4);
    \draw[very thick] (8,0) -- (0,4);
    \node[] at (-1,4) {$a = a_1$};
    \node[] at (9,4) {$a = a_0$};

    \draw[very thick, Blue] (0,1) -- (4,2) -- (8,1);

    \draw[very thick, JungleGreen] (0,0) -- (4,2) -- (8,0);
    \draw[line width=2.5pt, JungleGreen] (0,0) circle (0.1);
    \draw[line width=2.5pt, JungleGreen] (8,0) circle (0.1);
    \filldraw[JungleGreen] (0,1) circle (0.1);
    \filldraw[JungleGreen] (8,1) circle (0.1);

    \node[JungleGreen] at (-0.5,1) {$1/4$};
    
    \draw[very thick, dashed, Brown] (2,0) -- (2,1.5);
    \node[Brown] at (2,-0.5) {$\mu = 1/4$};
    \draw[very thick, dashed, Brown] (0,1.5) -- (2,1.5);
    \node[Brown] at (-0.5,1.5) {$3/8$};

    \node[draw, JungleGreen, fill=JungleGreen, shape=rectangle, anchor=center] (label1) at (9,2) {};
    \node[anchor=west, JungleGreen] at (label1.east) {{$\displaystyle \int u(m(\mu), t) d\mu$}};
    
    \node[draw, Blue, fill=Blue, shape=rectangle, anchor=center] (label2) at (9,1) {};
    \node[anchor=west, Blue] at (label2.east) {cav{$\displaystyle \left(\int u(m(\mu), t) d\mu\right)$}};
    
    \end{tikzpicture}
  \caption{Sender's Indirect Utility and Concavification}
  \label{B1_fig1}
\end{figure}

\cite{KamenicaGentzkow11} then imply the optimal experiment induces $\mu = \frac12$ with probability $\frac12$, and fully reveal state $0$ with probability $\frac12$. This guarantees a payoff of $\frac14(\frac12) + \frac12(\frac12) = \frac38$, as marked on the graph. 
The induced allocation rule is given by $p(t_0) = \frac12 a_0  + \frac12 a_1$ and $p(t_1) = a_1$, and so $u(p(t_0), t_0) = \frac12$ and $u(p(t_1), t_1) = 0$. 
Finally, to show $p$ is not implementable by a direct, belief-based mechanism, it is sufficient to show that no direct, belief-based mechanism of the form 
    \[ m'(\mu) = \begin{cases} p(\mu) & \text{   if  } \mu \in \{0, 1\}  \\ r(\mu) & \text{   else  } \end{cases} \]
    Sender's indirect utility (and concavification) from $m'$ are graphed in Figure \ref{B1_fig2} below. 
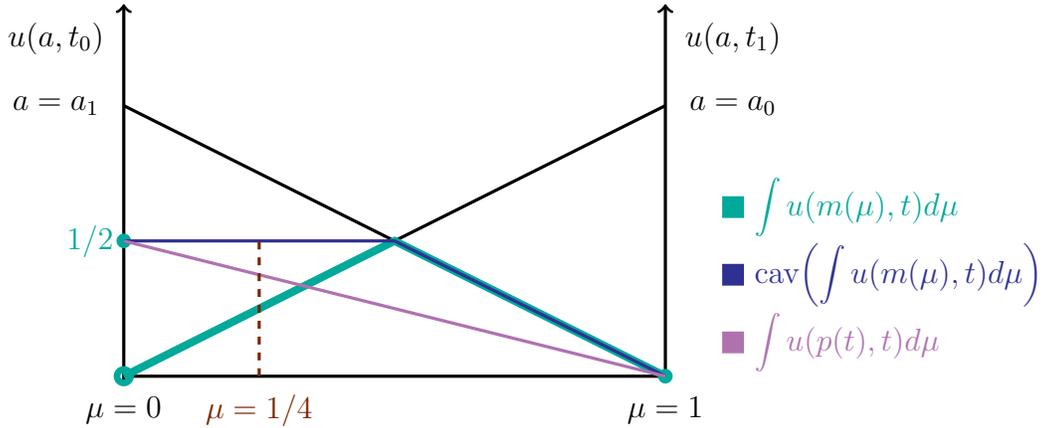
\begin{figure}[h]
\centering
    \begin{tikzpicture}[scale=.85]
    
    \draw[very thick] (0,0) -- (8,0);
    \draw[very thick, ->] (0,-0.1) -- (0,5.5);
    \draw[very thick, ->]  (8,-0.1) -- (8,5.5);
    \node[] at (0,-0.5) {$\mu = 0$};
    \node[] at (8,-0.5) {$\mu = 1$};
    \node[] at (-1,5) {$u(a,t_0)$};
    \node[] at (9,5) {$u(a,t_1)$};

    \draw[very thick] (0,0) -- (8,4);
    \draw[very thick] (8,0) -- (0,4);
    \node[] at (-1,4) {$a = a_1$};
    \node[] at (9,4) {$a = a_0$};

    \draw[line width=3pt, JungleGreen] (0,0) -- (4,2) -- (8,0);
    \draw[line width=2.5pt, JungleGreen] (0,0) circle (0.1);
    \filldraw[JungleGreen] (0,2) circle (0.1);
    \filldraw[JungleGreen] (8,0) circle (0.1);

    \draw[very thick, Blue] (0,2) -- (4,2) -- (8,0);

    \draw[very thick, Orchid] (0,2) -- (8,0);
    
    \node[JungleGreen] at (-0.5,2) {$1/2$};
    
    \draw[very thick, dashed, Brown] (2,0) -- (2,2);
    \node[Brown] at (2,-0.5) {$\mu = 1/4$};

    \node[draw, JungleGreen, fill=JungleGreen, shape=rectangle, anchor=center] (label1) at (9,2.5) {};
    \node[anchor=west, JungleGreen] at (label1.east) {{$\displaystyle \int u(m(\mu), t) d\mu$}};
    
    \node[draw, Blue, fill=Blue, shape=rectangle, anchor=center] (label2) at (9,1.5) {};
    \node[anchor=west, Blue] at (label2.east) {cav{$\displaystyle \left(\int u(m(\mu), t) d\mu\right)$}};

    \node[draw, Orchid, fill=Orchid, shape=rectangle, anchor=center] (label3) at (9,0.5) {};
    \node[anchor=west, Orchid] at (label3.east) {{$\displaystyle \int u(p(t), t) d\mu$}};
    
    \end{tikzpicture}
  \caption{Sender's New Indirect Utility and Concavification}
  \label{B1_fig2}
\end{figure}

    The optimal experiment from the indirect mechanism gives a strictly higher payoff than the fully revealing experiment, implying $p$ is not implemented by the direct mechanism. Alternatively, we can verify (graphically) that the linear functional defined by $\int u(p(t), t) d\mu$ crosses the grim-trigger correspondence from above, and as a result pointwise incentive constraints fail and Theorem \ref{t: grim trigger robustness principle} implies $p$ cannot be implemented by a direct belief-based mechanism. 
\end{proof}

Proposition \ref{behavioral_with_loss} contrasts with previous results in the contracting for experimentation literature (such as \cite{yoder2022designing}), who find belief-measurability of experiments is without loss when experimentation is costly, this cost is Sender's (known) private information, and Designer has access to transfers.
It also contrasts to \cite{dekel_etal}, who show restricting to the signal-choice model (which is analogized in the persuasion case as belief-based mechanisms) is without loss of generality; in our formalism, the ability to contract on the experiment is important. Second, 

\begin{proposition}
There exists a tuple $\{\mcal T, \mcal A, u, \mu^0\}$ such that at $\mu^0$, $\mcal P^B \subsetneq \mcal P^E$. 
\end{proposition}
\begin{proof}
    Consider again the setting in Proposition \ref{behavioral_with_loss}, and let $p$ be the (random) allocation rule considered in that example. We showed in that example that this allocation rule $p$ could not be implemented by a belief-based mechanism. However, it can be implemented at the experiment-based grim-trigger mechanism 
\[ m^{E}(\tau, \mu) = \begin{cases} p(t) & \text{ 
 if  } \tau = \tau^* \text{  and   } \mu = \delta_t \\ r(\mu) & \text{  else  } \end{cases} \]
 where $\tau^*$ is fully revealing. This is because pointwise incentive constraints hold at the prior, which is enough for experiment-implementability at a prior belief by Theorem \ref{t: implementability}.
\end{proof}

Finally, we show the key force driving Proposition \ref{behavioral_with_loss} is the desire to induce many actions which are below a payoff-guarantee. When restricting to mechanisms that obviate this behavior, direct implementation is without loss in the class of belief-based mechanisms. 

\begin{proposition}
\label{appendix deterministics}
   Suppose $p$ is an allocation rule such that either 
   \begin{enumerate}
       \item (Payoff Guarantee). $p: \mcal T \to \Delta(\mcal A)$ provides a \emph{payoff guarantee} where $u(p(t), t) \geq u(a, t)$ for some fixed $a$ at every $t$; or
       \item (Determinism). For every $t$, $p(t)$ supports at most one $a$. 
   \end{enumerate}
   Then any allocation rule $p$ which is implementable by a belief-based mechanism is implementable by a direct, belief-based, grim-trigger mechanism. 
\end{proposition}
\begin{proof}
For the first case, note that if $p$ provides a payoff guarantee, then 
\[ \int u(p(t), t) d\mu \geq \int u(r(\mu), t) d\mu \]
and so $p$ is implementable by a direct grim trigger mechanism by Theorem \ref{t: grim trigger robustness principle}.

The second case. 
If preferences satisfy homogeneous least favorites, then we are done. Else, fix a mechanism $m(\mu)$ inducing an allocation rule $p$ via experiment $\tau$.
Note that if $p$ is deterministic, then it must be that for any $\mu \neq \mu'$ such that $\mu$ and $\mu'$ are both supported by $\tau$, either $m(\mu) = m(\mu')$ or $\supp(\mu) \cap \supp(\mu') = \varnothing$, as otherwise there exists some $t \in \supp(\mu) \cap \supp(\mu')$ which receives actions $m(\mu) \neq m(\mu')$ both with positive probability, contradicting the hypothesis $p$ was deterministic. 
Thus, if $p$ is deterministic, then for any $(m, \tau)$ that induce $p$, the experiment $\tau$ can be thought of as a partition of the state space with respect to the prior $\mu$. 
Define the objects 
\[ \mathbf{U}(\mu) = \int u(m(\mu), t) d\mu \text{,   } \mathbf{U'}(\mu) = \int u(m'(\mu), t) d\mu, \text{  and  } C(\mu) = \int u(p(t), t) d\mu   \]
respectively, where $m'$ is the direct, grim trigger mechanism given by $m'(\mu) = p(t)$ if $\mu$ is degenerate at $t$, and $r(\mu)$ otherwise, 
where $r(\cdot)$ is a selection from the grim trigger correspondence. 
$\mathbf{U}(\mu)$ is Sender's indirect utility in the original mechanism, $\mathbf{U}'(\mu)$ is their indirect utility in the direct mechanism, and $C(\mu)$ is the payoff a Sender with belief $\mu$ obtains when they fully reveal according and obtain payoff $p(t)$ at each state. 
The concavification theorem of \cite{KamenicaGentzkow11} tells that the equilibrium experiment attains payoffs $\text{cav}(\mathbf{U}(\mu))$ and $\text{cav}(\mathbf{U}'(\mu))$. Given this result, it is sufficient to show 
\[ \text{cav}(\mathbf{U}'(\mu^0)) = C(\mu^0) = \text{cav}(\mathbf{U}(\mu^0))\]
for full-revelation to be an equilibrium under $m'$. The last equality is always true because $p(t)$ is induced by $m$. Moreover, 
$\text{cav}(\mathbf{U}'(\mu)) \geq C(\mu)$ since $C(\mu)$ is a concave function which can be induced by some experiment at any belief (particularly, complete full-revelation).
Finally, note that $\text{cav}(\mathbf{U}(\mu))$ is linear in the belief on the convex hull of $\supp(\tau)$ (denoted \text{co}$(\supp(\tau))$), since these are the beliefs induced in equilibrium. Moreover, because $m(\mu)$ is constant and gives action $p(t)$ if $t \in \supp(\mu)$, we know that it is linear and equal to $C(\mu)$ on \text{co}$(\supp(\tau))$. Moreover, on each element $\mu' \in \supp(\tau)$, because $p(t)$ is constant, we know that $\text{cav}(\mathbf{U}'(\mu')) = C(\mu')$. 
Taking these two facts together implies that for every belief $\tilde \mu \in \text{co}(\supp(\tau))$, it must be that 
\[ C(\tilde \mu)  = \text{cav}(\mathbf{U}(\tilde \mu)) \geq \text{cav}(\mathbf{U}'(\tilde \mu))  \]
Together, and using the fact $\mu^0 = \bb{E}_\tau[\mu']$ implies $\text{cav}(\mathbf{U}'(\mu^0)) = C(\mu^0)$. Thus, full revelation attains exactly the concavification value, as desired. 
\end{proof}
\end{document}